\documentclass{iopart}
\usepackage{dcolumn}
\usepackage{bm}
\usepackage{graphicx}
\usepackage{hyperref}
\usepackage{mathptmx}
\usepackage{subfigure}
\usepackage[utf8]{inputenc}
\usepackage{bbold}
\usepackage{color}
\usepackage{xcolor}
\usepackage{caption}

\newcommand{\floor}[1]{\left\lfloor #1 \right\rfloor}
\newcommand{\nint}[1]{\left\lfloor #1 \right\rceil}

\newcommand{\Epar}{E_{\|}}
\newcommand{\Eperp}{E_\bot}

\makeatletter

\renewcommand{\vec}[1]{\mathbf{#1}}

\newcommand{\eqref}[1]{(\ref{#1})}

\begin{document}

\title{Efficient algorithm for simulating particles in true quasiperiodic environments}

\author{Alan Rodrigo Mendoza Sosa and Atahualpa S.~Kraemer}

\address{
Departamento de F\'isica, Facultad de Ciencias, Universidad Nacional
Aut\'onoma de M\'exico,
Ciudad Universitaria, M\'exico D.F.\ 04510, Mexico
}

\eads{\mailto{alanmendoza@ciencias.unam.mx}
and 
  \mailto{ata.kraemer@ciencias.unam.mx}}

  \date{\today}

\begin{abstract}
We introduce an algorithm based on Generalized Dual Method (GDM) to efficiently study the dynamics of a particle in quasiperiodic environments without the need to use periodic approximations or to save the information of the vertices that make up the quasiperiodic lattice. 
We show that the computation time and the memory required to find the tile in which a particle is located as a function of the distance $R$ to the center of symmetry remains constant in our algorithm, while using the GDM directly both quantities go like $R^2$.
This allows us to perform realistic simulations with low consumption of computational resources. 
The algorithm can be used to study any quasiperiodic lattice that can be produced by the cut-and-project method. 
Using this algorithm, we have calculated the free path length distribution in quasiperiodic Lorentz gases reproducing previous results and for systems with high symmetries at the Boltzmann-Grad limit. 
We have found for the Boltzmann-Grad limit, that the distribution of free paths depends on the rank $r$ of the quasiperiodic system and not on its symmetry. 
%
%
The distribution as a function of the free path length $l$ appears to be a combination of exponential decay and a power-law behavior. The latter seems to become important only for probabilities less than $(2^{r-2} r (r+1))^{-1}$, showing an exponential decaying free-path length distribution for $r \rightarrow \infty$, similar to what is observed in disordered systems.
\end{abstract}

\maketitle
\section{Introduction}

Quasicrystals are materials that have a long-range order but lack translational symmetry; instead, the position of their atoms follows a quasiperiodic distribution \cite{Shechtman1984, Levine1984, Steinhardt1987, Senechal1996}.
These materials generally have symmetries ``forbidden'' by classical crystallographic theory, such as 5-, 8-, or 12-fold rotational symmetries \cite{Socolar1985, Steurer1990}.

Schechtman \cite{Shechtman1984} discovered these materials by abruptly reducing the temperature of an Aluminum-Manganese alloy trying to generate a metallic glass. 
The result was a crystalline phase with pentagonal symmetry, which he called the icosahedral phase. 
Following Schechtman’s work, several groups found similar structures that depended on the cooling rate of the alloys \cite{Schaefer1986, Nissen1988}, varying from the formation of a periodic crystal at slow rates of cooling, to the formation of metallic glass at high rates, passing through quasicrystals and crystalline structures called quasiperiodic approximants that locally had pentagonal symmetry at intermediate cooling rates \cite{Kelton2004}. 
Later stable quasicrystals were found \cite{Dubost1986, Ohashi1987, Tsai1988, tsai1989icosahedral, He1990} and in other kinds of systems \cite{Zeng2004, Zeng2005, Dotera2011}. 
More recently, quasicrystalline colloids \cite{Roichman05, Mikhael2008, Mikhael2010, Schmiedeberg2012, Martinsons2014} produced by laser interference have been studied, as well as metamaterials such as photonic \cite{Jin1999, Zoorob2000, Bayindir2001, Negro2003, Man2005, Lai06, Steurer2007, Barber2009, Jin2020, Fernandes2021} and phononic crystals \cite{He1989, Torres2003, Lai2005, Steurer2007, Zhang2007, SutterWidmer2007, Han2020}, in addition to the fact that quasiperiodic systems have been found at the rotated interface of graphene monolayers \cite{PhysRevB.99.165430, PhysRevLett.124.036803}. 
These discoveries and production techniques generated an interest in the study of quasiperiodic environments, such that in 2011 the Nobel Prize in Chemistry was given to Schechtman for the discovery of quasicrystals \cite{NobelPrizeSchechtman}.
Such interest continues to grow today, width many open problems from a mathematical point of view \cite{adiceam2016open}, and computational challenges, such as the statistical description of the quantum energy levels of a quasicrystal recently addressed with the use of random matrices \cite{grimm2021gaussian}, which also could be related with the length distribution of the free flights in a Lorentz gas \cite{Marklof2021QuantumOscilators}.

One of the reasons why it is interesting to study quasicrystals, is the flexibility they offer due to the variety of symmetries that can be produced, which can be of great importance in the optoelectronics industry \cite{Notomi2004, Feng2005, Hayat21}, making high symmetry quasicrystals attractive. 
Another reason why these systems are interesting is the existence of phasons \cite{Levine1985, Socolar1986, Boissieu2011, Kromer2012, Hielscher2017}, that is, thermal excitations that propagate in the lattice without energy cost generating phasonic flips \cite{hielscher2020phasonic}, which contribute to the specific heat in a similar way in which the phonons do in the periodic crystals. 
Just as phonons have a set of $d$ phononic modes, where $d$ is the dimension of the system, phasons also have a set of $r-d$ phasonic modes, where $r$ is the rank of the quasicrystal, i.e., the number of wave vectors linearly independent necessary to generate the reciprocal lattice of the quasiperiodic structure \cite{Lifshitz2007, Schmiedeberg2012, Martinsons2019}.
Due to this, we can conclude that if $r >> d$ any effect of the phasons will be dominant. 
For systems with rotational symmetry of order $N$, $r = \phi(N)$, where $\phi$ is the Euler’s totient function, which measures the number of primes relative to $N$ less than $N$. 
This implies that to have a higher value of $r$ is necessary to have a high symmetry, although an increase in symmetry does not necessarily increase the rank.

One of the problems of studying aperiodic systems numerically is precisely the lack of efficient algorithms that do not require excessive memory consumption without sacrificing the lack of periodicity. 
Usually, to solve this in quasiperiodic systems, a finite (albeit large) quasiperiodic array of vertices is produced via one of the three main methods: (i) substitution rules or deflation-inflation algorithm \cite{Gardner1977, Frank2008}, (ii) cut-and-project methods \cite{Elser1985, Senechal1996} and (iii) generalized dual method (GDM) \cite{Bruijn1981, Socolar1985, Naumis2003}. 
This array is generated at a particular size so that applying periodic boundary conditions minimizes the boundary effect by making the real vertices of the quasicrystal almost coincide with the vertices after applying the periodic condition \cite{kahlitz2012phase, schmiedeberg2008colloidal}. 
However, for many effects observed and predicted in quasiperiodic systems, it is still necessary to work with very large systems, for example, when calculating the band spectrum \cite{benza1991band}. In addition, as the rank of the system increases, the effects of periodic boundaries are more noticeable \cite{schmiedeberg2008colloidal}. 

Another example of this is the diffusion of particles in quasiperiodic Lorentz gases at the Boltzmann-Grad limit \cite{kraemer2015efficient, marklof2014free, marklof2015visibility}. 
In the case of a periodic Lorentz gas, the free path distribution goes like a power-law \cite{marklof2011periodic}, while a Poissonian distribution of obstacles produces an exponential decay \cite{wennberg2012free, marklof2011periodic}. 
The question is if quasiperiodic Lorentz gases are closer to the periodic cases or the disordered cases. 
For 1D it has been shown numerically that such distribution is a power-law with the same exponent as the periodic case. 
However, it has not been tested for higher dimensions, and the reason is the lack of an efficient algorithm as it has been recognized before \cite{wennberg2012free}. 
For 2D systems, as far as we know, only have been studied those quasiperiodic arrays that are the result of projecting the vertices of a 3D lattice onto a 2D plane and with positive obstacle radius, \cite{kraemer2013embedding, kraemer2015horizons}, so, they have no rotational symmetry, there are channels (where particles can move infinitely without colliding), and they are not in the Boltzmann-Grads limit. 
Following the same technique as in \cite{kraemer2013embedding},  it is possible in principle to work with lattices of higher dimension (and where quasicrystals with rotational symmetry could be embedded). 
However, the numerical complications of calculating the convex hull make this unfeasible, except for low dimensions (ranks) \cite{chazelle1993optimal}. 
Another approach is to study Lorentz gases formed by several overlapped periodic lattices with different orientations \cite{marklof2014power}. 
In this case, the obtained free path distribution is a power-law, where the exponent is $-n-2$, where $n$ is the number of overlapping lattices.
One could then conjecture that the quasiperiodic Lorentz gases of symmetry $N$ would generate a distribution of free paths as a power-law, whose exponent depends on $N$. 
However, as we will show numerically, the quasiperiodic case is closer to the periodic case for low ranks, and closer to the Poissonian process when the rank is high. 

In this paper, we first introduce an algorithm based on the GDM to generate the locale around any point on a quasiperiodic lattice with any symmetry. 
This allows us to simulate particles in quasiperiodic environments without the need to store an enormous amount of data in memory, which makes the algorithm efficient in both, memory consumption and computing speed. 
Then, we test the algorithm by obtaining the free path length distribution in periodic and quasiperiodic Lorentz gases first reproducing the results in \cite{kraemer2015horizons} and then studying the Boltzmann-Grad limit for various symmetries, showing that said distributions depend on the rank of the quasicrystal and not on the symmetry itself.

\section{Analytical expressions for vertices of a quasiperiodic lattice}

From the GDM it is possible to derive an analytical expression for the coordinates of the vertices of a quasiperiodic lattice \cite{Naumis2003}. 
In this section we will develop the algebra corresponding to the two-dimensional case, following the indications contained in the article by Naumis and Aragón \footnote{In this article equations 8 and 10 have an error in one of their signs.}, considering only the quasiperiodic arrangements built from a grid with a constant separation between its lines, which correspond to those that can be produced via the cut-and-project method.

Let $\mathbf{S} = \{ \vec{e}_{1}, \vec{e}_{2}, \dots, \vec{e}_{N} \ | \ \vec{e}_{i} \in \mathbb{R}^{2}, \ N \in \mathbb{Z}^{+} \}$ be a set of arbitrary vectors (not necessarily unitary), called star vectors, which determine the rotational order of the quasiperiodic lattice to be constructed. 
We define the grid generated from the star vectors as 
\begin{equation*}
    \mathbf{G}_{N} = \{ \vec{x} \in \mathbb{R}^{2} \ | \ \vec{x} \cdot \vec{e}_{i} = n_{i} + \alpha_{i}; \ i \in \{1, 2, \dots, N\}, \ n_{i} \in \mathbb{Z}\},
\end{equation*}
where $\alpha_{i} \in (0, 1)$ is a displacement with respect to the origin of the set of lines orthogonal to the vector $\vec{e}_{i}$. 
The interval for the $\alpha_{i}$ values is defined as open and not closed because one of the necessary conditions for the present algorithm requires that three or more lines of the grid $\mathbf{G}_{N}$ do not intersect at the same point, which occurs at $\alpha_{i} = 0$ or $\alpha_{i} = 1$ for all $i \in \{1, 2, \dots, N\}$.

Let $\vec{x}_{I}$ be the point of intersection of two arbitrary lines of the grid $\mathbf{G}_{N}$ associated with the star vectors $\vec{e}_{j}$ and $\vec{e}_{k}$. 
The coordinates of this point are determined by the solution to the following system of equations 
\begin{equation*}
    \vec{x}_{I} \cdot \vec{e}_{j} = n_{j} + \alpha_{j}, \ \ \vec{x}_{I} \cdot \vec{e}_{k} = n_{k} + \alpha_{k},
\end{equation*}
which can be written in matrix form as
\begin{equation*}
        \left( \matrix{ e_{jx} & e_{jy} \cr
e_{kx} & e_{ky} \cr}
 \right) \left( \matrix{x \cr
y \cr}
 \right)
 = \left( \matrix{
 n_{j} + \alpha_{j}\cr
n_{k} + \alpha_{k} \cr
}\right)
\end{equation*}
solving for the vector $ (x, y) $ we have
\begin{equation*}
\left( \matrix{
x\cr
y\cr
}
\right)
=
\frac{1}{e_{jx} e_{ky} - e_{jy} e_{kx}}
 \left( \matrix{
e_{ky} & -e_{jy}\cr
-e_{kx} & e_{jx}\cr
}
\right)
\left( \matrix{
n_{j} + \alpha_{j}\cr
n_{k} + \alpha_{k}\cr
}
\right)
\end{equation*}
\begin{equation*}
=
\frac{1}{A_{jk}}
\left( \matrix{
e_{ky} \left( n_{j} + \alpha_{j} \right) - e_{jy} \left( n_{k} + \alpha_{k} \right)\cr
-e_{kx} \left( n_{j} + \alpha_{j} \right) + e_{jx} \left( n_{k} + \alpha_{k} \right)\cr
}
\right),
\end{equation*}
where $A_{jk} = e_{jx} e_{ky} - e_{jy} e_{kx}$ is the area of the parallelogram generated by the vectors $\vec{e}_{j}$ and $\vec{e}_{k}$.

Given the vector $\vec{e}_{i} = \left(e_{ix}, e_{iy} \right)$ we define its orthogonal vector as $\vec{e}_{i}^{\perp} = \left(e_{iy}, -e_{ix} \right)$. 
Using this notation, we can write the point of intersection $\vec{x}_{I}$ as
\begin{equation}
    \vec{x}_{I} = \frac{1}{A_{jk}} \left[ \left( n_{j} + \alpha_{j} \right) \vec{e}_{k}^{\perp} - \left( n_{k} + \alpha_{k} \right) \vec{e}_{j}^{\perp} \right].
    \label{Intersection_Solution}
\end{equation}

\begin{figure}[ht]
    \centering
    \includegraphics[width=260pt]{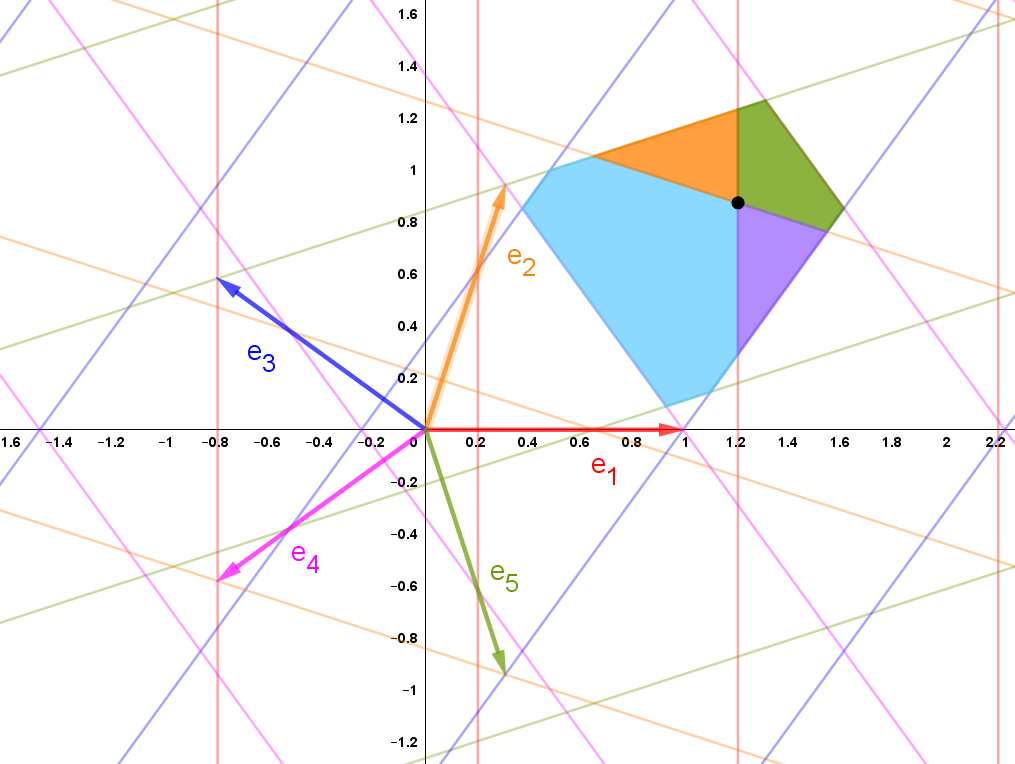}
    \captionsetup{justification=justified}
    \caption{(Color online) Grid $G_{N}$ generated by a set of star vectors corresponding to the vertices of a regular pentagon and constants $\alpha_{i} = 0.2 \ \forall i \in \{1, \dots, 5\}$. The black point corresponds to the intersection of the lines given by $\vec{x} \cdot \vec{e}_{1} = 1.2$ and $\vec{x} \cdot \vec{e}_{2} = 1.2$. In solid colors the four open regions associated with the intersection point.}
    \label{fig:Four_Sections_Grid}
\end{figure}

Under the assumption that only two lines of the grid $\mathbf{G}_{N}$ pass-through the point $\vec{x}_{I}$, it follows that around it there are four open regions delimited by the lines that make up the grid $\mathbf{G}_{N}$; see figure \ref{fig:Four_Sections_Grid}. 
These regions can include their border without affecting the generalized dual method (which consists of mapping regions to points), however they are considered open regions to avoid the ambiguity caused by closed regions with respect to their common borders.

The vertices of the quasiperiodic lattice are given by the dual transformation that maps each of these regions to the point given by the expression
\begin{equation}
    \vec{t} = \sum_{i = 1}^{N} m_{i} \vec{e}_{i},
    \label{ec:Mapeo_Dual}
\end{equation}
where $m_{i}$ is the minimum between the integers associated with the lines orthogonal to the vector $\vec{e}_{i}$ between which the region in question is contained.

For one of these regions (in the example in figure \ref{fig:Four_Sections_Grid}, the region in green), the integers $m_{i}$ can be calculated by projecting the intersection point $\vec{x}_{I}$ with the respective star vector $\vec{e}_{i}$, subtracting the parameter $\alpha_{i}$ from the result and taking the nearest integer below that value. 
Substituting in the equation \ref{ec:Mapeo_Dual} the integers $m_{i}$ by the expression to calculate them, we have that one of the vertices of the quasiperiodic lattice is given by

\begin{equation*}
\vec{t}_{n_{j}, n_{k}}^{\ 0} = n_{j} \vec{e}_{j} + n_{k} \vec{e}_{k} 
\end{equation*}
\begin{equation}
  + \sum_{i \neq j \neq k}^{N} \floor{\frac{1}{A_{jk}} \left[ \left( n_{j} + \alpha_{j} \right) \vec{e}_{k}^{\perp} - \left( n_{k} + \alpha_{k} \right) \vec{e}_{j}^{\perp} \right] \cdot \vec{e}_{i} - \alpha_{i}} \vec{e}_{i}.
   \label{ec:Punto_1}
\end{equation}

The other three points associated with the three adjacent regions are given by
\begin{equation}
   \vec{t}_{n_{j}, n_{k}}^{\ 1} =  \vec{t}_{n_{j}, n_{k}}^{\ 0} - \vec{e}_{j},
   \label{ec:Punto_2}
\end{equation}
\begin{equation}
    \vec{t}_{n_{j}, n_{k}}^{\ 2} =  \vec{t}_{n_{j}, n_{k}}^{\ 0} - \vec{e}_{j} - \vec{e}_{k},
    \label{ec:Punto_3}
\end{equation}
\begin{equation}
    \vec{t}_{n_{j}, n_{k}}^{\ 3} =  \vec{t}_{n_{j}, n_{k}}^{\ 0} - \vec{e}_{k}.
    \label{ec:Punto_4}
\end{equation}

\section{Neighborhood of the quasiperiodic lattice around an arbitrary point}

\begin{figure}
    \centering
    \includegraphics[width=260pt]{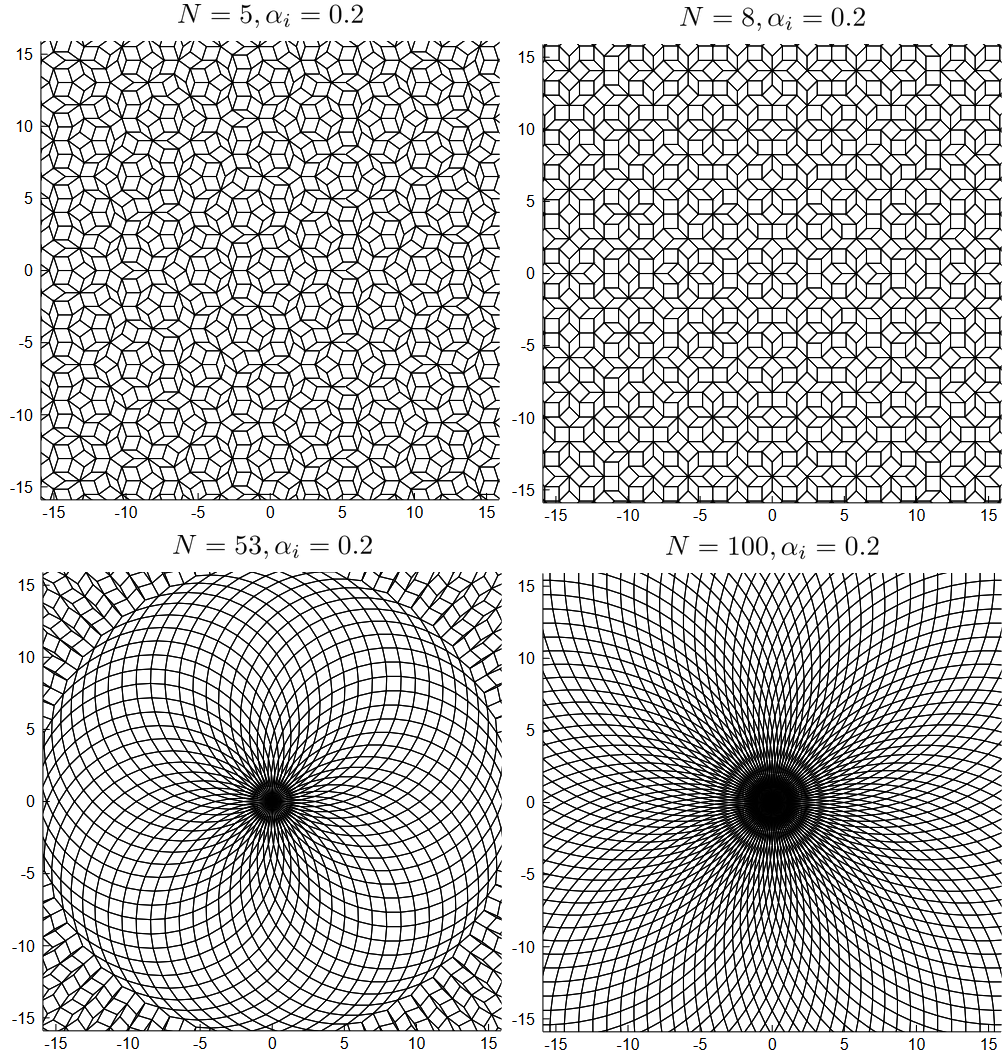}
    \caption{Examples of quasiperiodic lattices generated from the equations \ref{ec:Punto_1} - \ref{ec:Punto_4} when considering all pairs of non-collinear star vectors $(\vec{e}_{j}, \vec{e}_{k})$ as well as all pairs of integers $\left(n_{j}, n_{k} \right)$ where $n_{j}, n_{k} \in [-10,10]$. The integer $N$ corresponds to the number of sides of the regular polygon inscribed in the unit circle that defines the set of star vectors. In all cases, the parameter $\alpha_{i} = 0.2 \ \forall i \in \{ 1, 2, \dots, N \}$. The axes shown in the images correspond to the X and Y positions in real space.}
    \label{fig:Examples_Lattices}
\end{figure}

The equations \ref{ec:Punto_1} - \ref{ec:Punto_4} give us the vertices of any of the tiles that make up the quasiperiodic lattice knowing the integers $n_{j}$ and $n_{k}$ associated with some pair of the star vectors $\vec{e}_{j}$ and $\vec{e}_{k}$. 
Due to this, we would like, given any point in the plane, to know which integers and which star vectors generate the tile that contains said point.
A naive way to do this is to consider all combinations of integer pairs into a range of values $[-c,c]$, as well as all combinations of star vectors until find which tile contains the point of interest. 
This method can be functional to construct quasiperiodic lattices around the center of symmetry (see figure \ref{fig:Examples_Lattices}), but since the number of tiles grows as $R^2$, where $R$ is the distance from the point to the center of symmetry, the computational complexity to find the right pair of numbers and vectors grows at least as $R^2$.

\begin{figure}[ht]
    \centering
    \includegraphics[width=260pt]{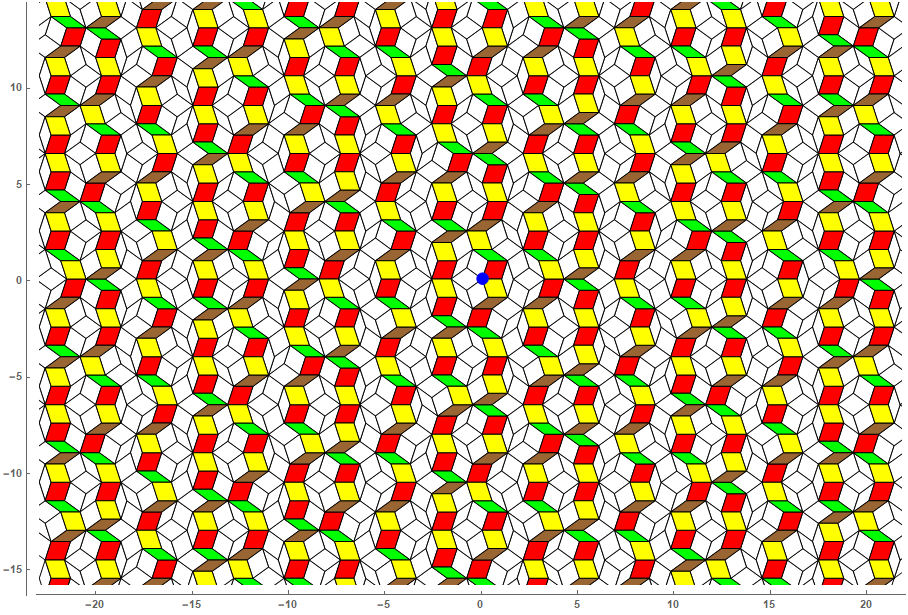}
    \caption{(Color online) Quasiperiodic lattice with $N = 5$ centered on the origin. The combination of star vectors $\left( \vec{e}_{j}, \vec{e}_{k} \right)$ that gives rise to the colored tiles is as follows: red - $\left( \vec{e}_{1}, \vec{e}_{2} \right)$, green - $\left( \vec{e}_{1}, \vec{e}_{3} \right)$, brown - $\left( \vec{e}_{1}, \vec{e}_{4} \right)$ and yellow - $\left( \vec{e}_{1}, \vec{e}_{5} \right)$ being the vector $\vec{e}_{1} = (1,0)$, enumerating the rest of the vectors counterclockwise. The bands appear in the order of the integer $n_{1}$, that is, the band closest to the origin (indicated by a blue dot) on the right corresponds to the integer $n_{1} = 0$, the next band to the right corresponds to the integer $n_{1} = 1$ and so on (analogously for the bands on the left). The axes shown in the images correspond to the X and Y positions in real space.}
    \label{fig:Stripes}
\end{figure}

To improve on this naive algorithm, let us observe that when generating the lattices centered on the origin using this algorithm, the tiles that are constructed by fixing one of the star vectors are grouped forming \emph{bands} orthogonal to it (see figure \ref{fig:Stripes}), each one associated with an integer. 
Therefore, if there is an average in the separation of the bands and the maximum separation has an upper bound, we can limit the set of integers that we must review regardless of the position of the point, which allows the complexity of the algorithm to find the tile that contains the point to become independent of $R$.

The set of all vertices of a quasiperiodic lattice that can be generated with the cut-and-project method can be interpreted as an $N-2$ dimensional object that passes through a hypercube of dimension $N$ \cite{kraemer2013embedding} (which in turn is equal to the number of star vectors used in the GDM) which we will call \emph{acceptance domain}. 
Changing the integer $n_i$ means moving to some neighboring hypercube and the distance between the acceptance domains will be limited by some value that depends on the inclination of this domain and the dimension $N$ of the hypercube. 
We also know that the average distance is well defined since the acceptance domain is finite. 
Therefore we expect that the distance between the bands has a well-defined average that depends on of star vectors (the symmetry of the system), and the maximum distance between them is bounded.

To approximate the value of $n_i$ associated with the star vector $\vec{e}_{i}$ in such a way that the generated band passes through the point $\vec{P}$ or close to it, we simply have to project $\vec{P}$ into the unitary vector $\hat{\vec{e}}_{i}$ and divide the distance from the origin to the projection by the average distance between the bands. 
Since this is not necessarily an integer, we must also apply the nearest integer function to get $n_i$:

\begin{equation}
    n_{i} = \nint{\frac{ \vec{P} \cdot \hat{\vec{e}}_{i}}{d_{A}} },
    \label{ec:Aproximacion_Enteros}
\end{equation}
where $d_{A}$ is the average distance between the bands and $\nint{\cdot}$ is the nearest integer.

For the particular case in which the quasiperiodic lattice is symmetric and unitary, i.e. the case in which the star vectors that generate the quasiperiodic lattice correspond to the vertices of a regular $N$-sided polygon inscribed in the unit circle, the average distance $d_{A}$ between bands is equal to $N/2$. Throughout this work, unless otherwise indicated, we will be considering only symmetric and unitary quasiperiodic lattices.

\begin{figure}[ht]
    \centering
    \includegraphics[width=300pt]{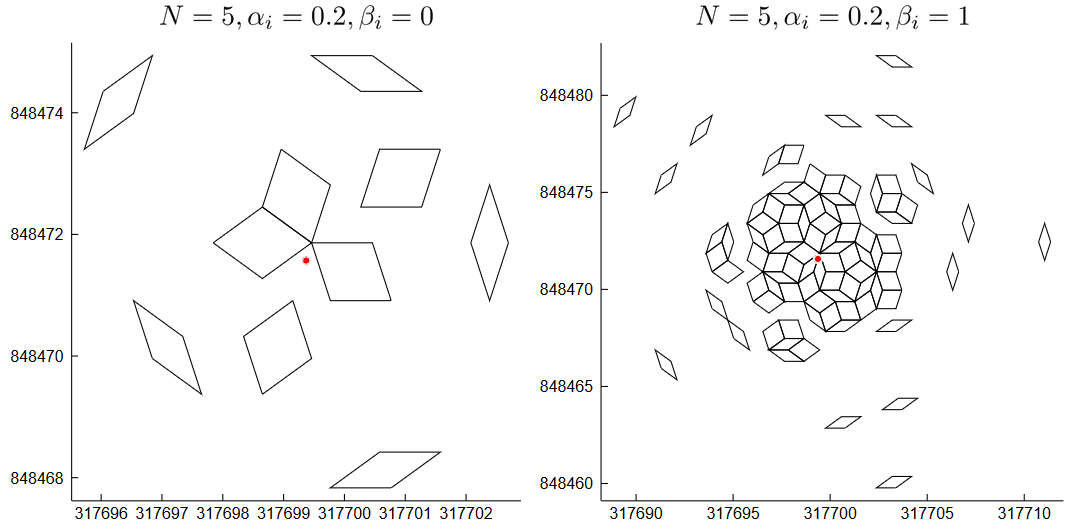}
    \caption{(Color online) Neighborhood of a quasiperiodic lattice with rotational symmetry $N = 5$ generated around the red point. In the image on the left we have the tiles generated without considering margin of error ($\beta_{i} = 0 \ \forall i \in \{ 1, 2, \dots, N \}$). In the image on the right we have the tiles generated considering a margin of error $\beta_{i} = 1 \ \forall i \in \{ 1, 2, \dots, N \}$. The axes shown in the images correspond to the X and Y positions in real space.}
    \label{fig:Vecindades_Margen_Error}
\end{figure}

Note that the first band in a generic way does not pass through the origin and the bands are not straight, so it may be necessary to consider a margin of error for each of the integers obtained by this method (see figure \ref{fig:Vecindades_Margen_Error}). 
In other words, when generating the tile associated with the star vector pair $\left( \vec{e}_{i}, \vec{e}_{k} \right)$, we must consider the pairs of integers $\left(d_{i}, d_{k} \right)$ with $d_{l} \in \left[ n_{l} - \beta_{l}, n_{l} + \beta_{l} \right]$ where $\beta_{l} \in \mathbb{Z}^{+} \cup \{0\}$ is associated with the margin of error considered for the integer $n_{l}$. In general, the parameters $\beta_{l}$ determine the size of the neighborhood of the quasiperiodic lattice generated around a point $\vec{P}$: the larger the value of $\beta_{l}$, the larger the size of the neighborhood of the quasiperiodic lattice. 

\subsection{Main cluster}
Using equations \ref{ec:Punto_1} - \ref{ec:Punto_4} for all pairs of star vectors with their corresponding integers $n_i$, different tiles are generated around the desired point in such a way that they adjoin each other. 
We call the set of these tiles the main cluster. 
However, some tiles are also produced that are not part of this main cluster as shown in figure \ref{fig:Vecindades_Margen_Error}. 
When simulating the dynamics of a particle in our quasiperiodic environment we want to take advantage of the already built tessellation. 
For this we store the vertices of the tiles of the main cluster in the memory and calculate the trajectory of the particle until it is no longer contained in the main cluster, after which we clean the memory and generate a new neighborhood around the last known position of the particle. 
Due to the above, the tiles that do not belong to the main cluster are not of interest and we must eliminate them so as not to consume computational resources unnecessarily. 

\begin{figure}
    \centering
    \includegraphics[width=350pt]{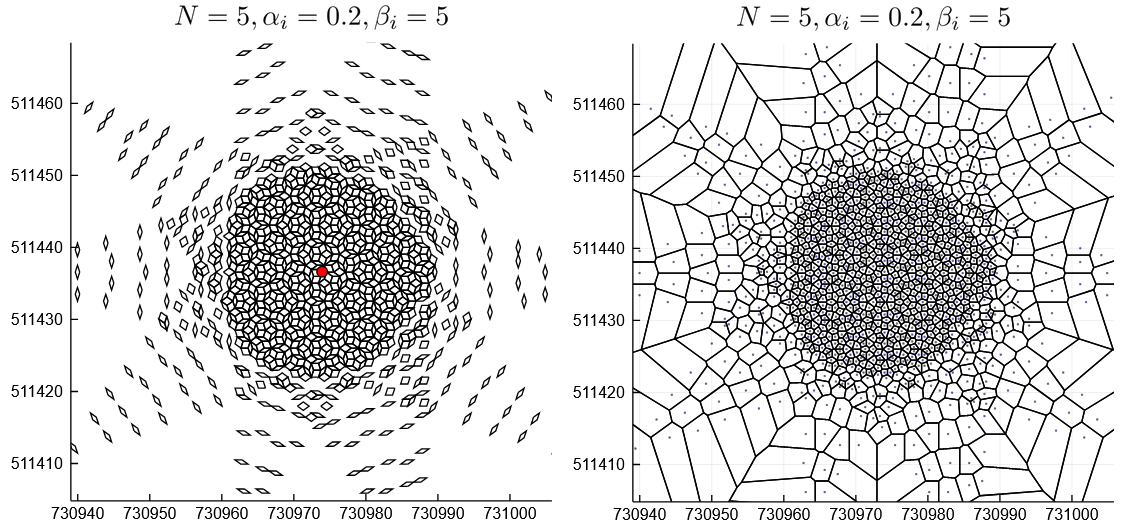}
    \caption{(Color online) Comparison between the neighborhood of a quasiperiodic lattice (left image) and the Voronoi polygons generated by taking as Voronoi centers the centers of the tiles that make up said lattice (right image). In red, the point around which the quasiperiodic lattice was built is indicated, while the centers of the tiles are indicated as small black points in the right image. The axes shown in the images correspond to the X and Y positions in real space.}
    \label{fig:Areas_Voronoi}
\end{figure}

\begin{figure}
    \centering
    \includegraphics[width=350pt]{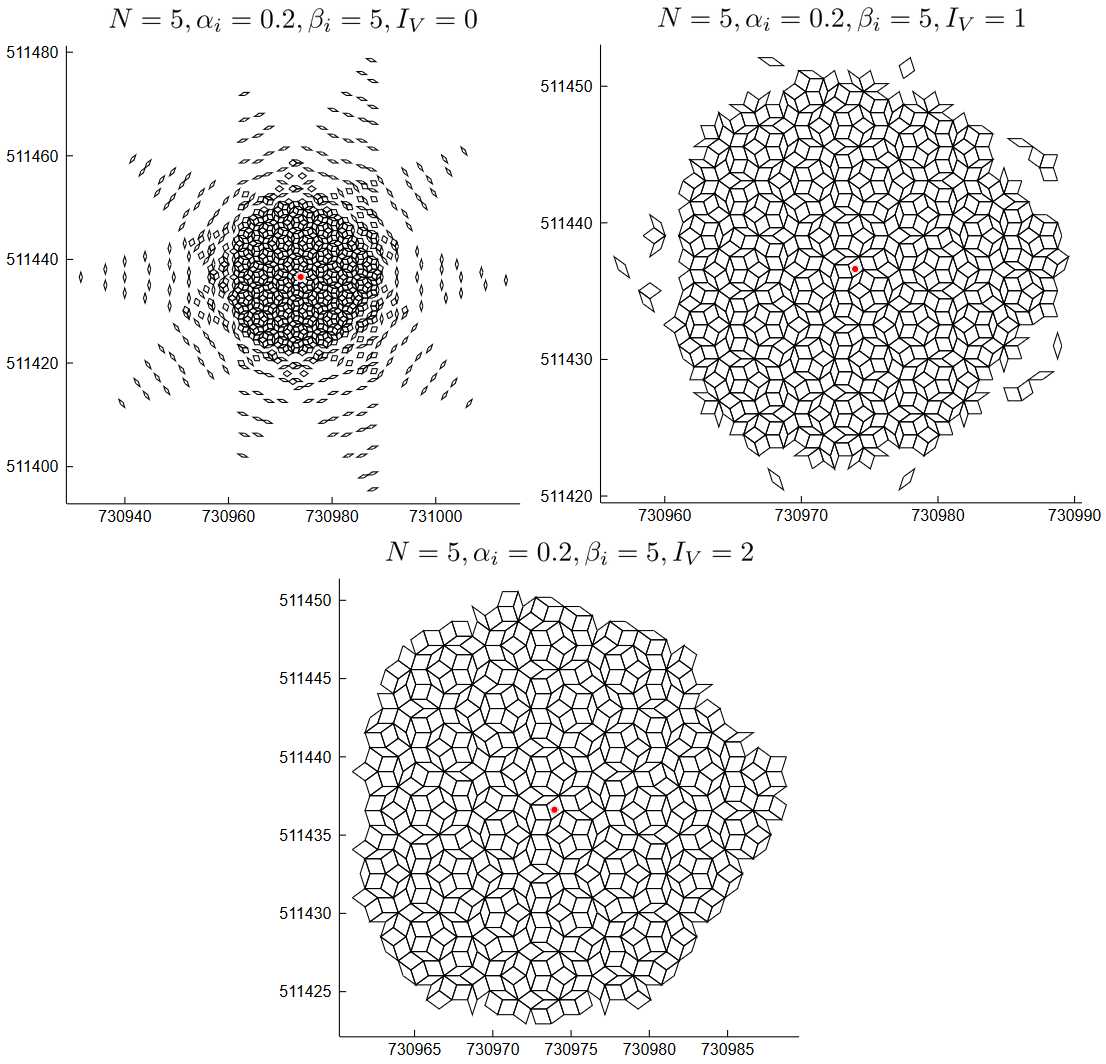}
    \caption{(Color online) The neighborhood of a quasiperiodic lattice with rotational symmetry $N = 5$ is built around the red point. The margin of error used is $\beta_{i} = 5 \ \forall i \in {1, \dots, N}$. The parameter $I_{V}$ is the number of iterations of the algorithm to preserve only the main cluster of tiles using Voronoi tessellations. Note that for this particular case, two iterations are enough to keep only the main cluster. The numerical value of the upper bound for the area of the Voronoi polygons used is $A = 1.2$. The axes shown in the images correspond to the X and Y positions in real space.}
    \label{fig:Iteraciones_Voronoi}
\end{figure}

To eliminate the tiles that do not belong to the main cluster we propose to build a Voronoi tessellation on the centers of the tiles, such that the tiles that are on the border of the clusters, as well as the isolated tiles, will have associated polygons of Voronoi with areas greater than an upper bound $A$, which we will proceed to eliminate (see figure \ref{fig:Areas_Voronoi}). 
By iterating this algorithm, the isolated tiles are eliminated and the size of the tile clusters is reduced in such a way that, eventually, only the tiles that make up the main cluster are preserved (see figure \ref{fig:Iteraciones_Voronoi}).

\subsection{Circular neighborhood}

Once we have the main cluster where we ensure that the particle is located, we can simulate the movement of the particle until it reaches the frontier of the main cluster, in which case a new neighborhood of the quasiperiodic lattice is generated around the last known position of the particle. 
However, the computational time it takes to find when the particle is on the boundary generally depends on the number of tiles on the boundary. 
To reduce this computation time we can take advantage of the shape of the main cluster. 

One of the characteristics that the main cluster presents after applying the algorithm described in the previous section is that it can be approximated by a circle with center at point $\vec{P}$ from which the neighborhood of the quasiperiodic lattice is generated, whose radius depends on the symmetry rotation of the lattice as well as the values of the parameters $I_{V}, \alpha_{i}$ and $\beta_{i}$ with $i \in \{1, \dots, N\}$, where $I_{V}$ is the number of iterations of the algorithm to preserve only the main cluster of tiles using Voronoi tessellations. 
Thus, an alternative is to obtain a radius $R_{C}$ such that the circle centered at $\vec{P}$ and of radius $R_{C}$ is completely tessellated. In this case, we can consider that if the distance between the position of our particle and $\vec{P}$ is greater than $R_{C}$, then the particle is outside the main cluster.   

Given the rotational symmetry $N$ of the quasiperiodic lattice and setting the parameters $\alpha_{i}, \beta_{i} = \beta \ \forall i \in \{1, \dots, N\}$, as well as the number of iterations $I_{V} \in \mathbb{Z}^{+}$ of the algorithm to eliminate isolated tiles, the radius of the circle ($R_{C}$) that approaches the main cluster is obtained from the following algorithm:

\begin{enumerate}
    \item An arbitrary point $\vec{P}$ is randomly generated in the plane. Around point $\vec{P}$, a neighborhood of the quasiperiodic lattice is generated with the previously set parameters, of this neighborhood only the main cluster is preserved.
    
    \item The following values are calculated 
    \begin{equation*}
        |\textit{Max}_{x} - \vec{P}_{x}|, |\textit{Min}_{x} - \vec{P}_{x}|, |\textit{Max}_{y} - \vec{P}_{y}|, |\textit{Min}_{y} - \vec{P}_{y}|,
    \end{equation*}
    where $\textit{Max}_{x (y)}$ is the maximum value of the $x (y)$ coordinate of all the vertices of the main cluster and $\textit{Min}_{x (y)}$ is the minimum value of the $x (y)$ coordinate of those vertices. From this set of four values the smallest is selected.
    
    \item The previous steps are iterated until a good statistical sample is obtained and the average of the values obtained in step two is calculated. This average corresponds to the radius $R_{C}$.
\end{enumerate}

\begin{figure}
    \centering
    \includegraphics[width=220pt]{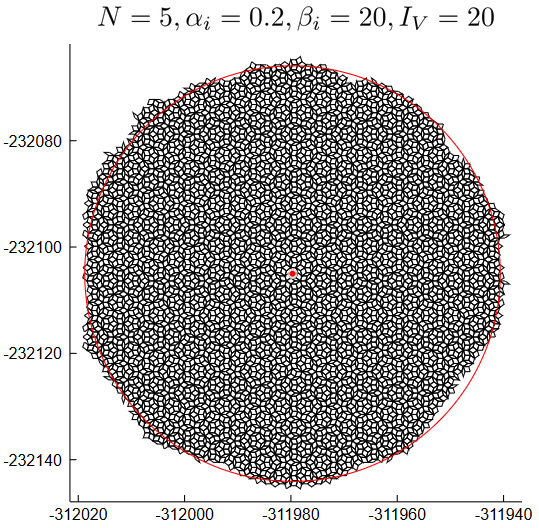}
    \caption{(Color online) Neighborhood of a quasiperiodic lattice with rotational symmetry $N = 5$ around the red point. The neighborhood was generated with a margin of error $\beta_{i} = 20 \ \forall i \in \{1, \dots, N\}$ and a parameter $I_{V} = 20$. To get the radius of the circle, $R_{C} = 39.08$ units, an average was carried out on $50$ neighborhoods. The axes shown in the images correspond to the X and Y positions in real space.}
    \label{fig:Vecindad_Circular}
\end{figure}

In general, the circle that approaches the main cluster of the quasiperiodic lattice presents blank regions in which there are no tiles (see bottom right of figure \ref{fig:Vecindad_Circular}).
In principle, when working with a circular neighborhood of the quasiperiodic lattice, we would like to be certain that the entire circle is filled by the tiles that make up the main cluster, for this purpose we define a parameter $\gamma \in (0, 1]$ that scales the value of the radius $R_{C}$ so that the circles with the scaled radius fulfill this property, that is:
\begin{equation*}
    R_{S} = \gamma \ R_{C},
\end{equation*}
where $R_{S}$ is a \emph{safe} radius.

\subsection{Choice of parameters}

Throughout this work we have defined some parameters required by our local algorithm to be able to generate a tessellation in a circular neighborhood around an arbitrary point in the plane. 
How to choose the parameters is still obscure. 
In this subsection we address the role of these and some useful bounds to choose them appropriately.

The first array of parameters is $\alpha_{i}$, which correspond to a displacement of the set of orthogonal lines with respect to the origin in the direction of the star vector $\vec{e}_{i}$ (before applying the linear transformation). 
The values that each $\alpha$ can take are any real number in the interval $(0,1)$. 
In general, for quasiperiodic arrays, given any set of values of $\alpha$, only 2 lines intersect at each point. 
The probability that there are more is $0$, although it is not impossible. 
For example, if all $\alpha$ were $0$, more than two lines intersect at the origin. 
If more than 2 lines intersect at the same point, this is equivalent to the parallel subspace $\Epar$ (from the cut-and-project method) passing through the origin. 
Since this subspace is a completely orthogonal space, this can only happen once. 
So, even if the choice of $\alpha$’s is unfortunate, it will only throw an error at one point in the plane, and that error can be corrected by substituting the set of tiles overlapped by a single tile which is the convex hull of the overlapping tiles. This parameter allows producing different quasiperiodic lattices associated with the same rotational symmetry, so its manipulation can be relevant to study issues related to phasonic flips, for example.
In our simulations we use $\alpha = 0.2$ but there is no special reason. We did tests with different values without ever detecting a problem.

The second set of parameters that we use in the algorithm are the $\beta_i$.
By computing the set of values $n_i$ with the equation \ref{ec:Aproximacion_Enteros}, a close neighborhood can be generated, but one that does not actually include the point of interest. 
So instead of fixing a single value for $n_{i}$, one can use values around $n_{i}$. How many more values we take into account is given by $\beta_{i}$. 
In the measurements we made, $\beta_{i} = 1 \ \forall i \in \{1, 2, \dots, N\}$ from N = 5 to N = 39, and $\beta_{i} = 0$ for all other symmetries. 
Our conjecture is that $\beta_i \leq 1$ always holds, although there might be special cases where this fails. 
In principle one can choose any larger value of $\beta$, which generates a larger neighborhood, but that also increases computation time, so the smaller the value of $\beta_{i}$, the more efficient the code. It is important to mention that the area of the main cluster grows as a function of $\beta$ but as a function of $N$. The number of rhombuses produced by the GDM equals the intersection of the lines in the dual grid. This leads to a number of tiles as a function of $\beta$ and $N$ as $ (1+2\beta)^2 N (N-1)/2$. Then, even for $\beta = 0$, the number of rhombuses grows fast as a function of $N$. Consequently, for large $N$, the value of $\beta$ must be equal or close to $0$, or memory overflows, but it is not necessarily a larger $\beta$. 

The third parameter is $A$. 
This parameter is used to remove the tiles that are disconnected from the main cluster and corresponds to an upper bound for the area of the Voronoi cells generated with the centroids of the tiles of the quasiperiodic array. 
The lower the value of this parameter, with fewer iterations it will be possible to remove all the vertices that do not correspond to the main cluster. 
If we have a Voronoi tessellation and add a center, this will change some of the Voronoi polygons, reducing their area. 
Therefore, the largest area of a Voronoi polygon that is surrounded by centers will be when there are the fewest centers around it. 
Each rhombus $R_{i}$ has 4 more rhombuses as neighbors and 4 rhombuses that coincide at the corners. 
The centers of these 8 rhombuses will be the potential neighbors of the center $\vec{p}$ of $R_{i}$. 
The four neighboring rhombuses of $R_{i}$ will have centers that are neighbors of $\vec{p}$ in the sense that the Voronoi polygon of each of them will share an edge with the polygon of $\vec{p}$. 
The minimum number of sides that the Voronoi polygon of $\vec{p}$ will have is when it involves only squares and the polygon will be a square of area $1$.
If we deform the squares, polygons of between $6$ and $8$ sides can appear. 
If we only consider $6$-sided polygons whose vertices are the $6$ centers closest to $\vec{p}$, the one with maximum area is a regular whose area is $1.5$ if the side of each rhombus is of length $1$, that is, if the star vectors are normalized. 
Therefore, any value greater than $1.5$ is a safe upper bound of $A$. 
This bound can probably be improved by considering the $8$ vertices, since the two remaining vertices must reduce the size of the bound of $A$ and in fact, by doing numerical tests we observe that $A = 1.2$ did not generate errors in any case.

The fourth parameter is $I_{V}$ and corresponds to the number of iterations required to remove the vertices that do not correspond to the main cluster. 
A high number of iterations can lead to eliminating so many vertices that the point around which we want to build the tessellation is outside the main cluster. 
On the other hand, a low number of iterations can leave tiles that do not correspond to the main cluster. 
This is the only parameter that we do not have a systematic way to obtain and whose variation was between $I_{V} = 2$ and $I_{V} = 18$. 
Typically for low symmetries the number of iterations is low, while for high symmetries it is high. 
To obtain the necessary number of iterations we did several trials and checked when there were no errors. 
In general, it would be best to take the smallest value of $I_{V}$ that ensures that only those vertices that do not belong to the main cluster are eliminated.

Finally, the last parameter $\gamma \in (0,1]$ has a double purpose: on the one hand only vertices that are inside a circle are chosen, on the other, if the iterations $I_{V}$ were not enough, they eliminate possible vertices that did not belong to the main cluster. 
This parameter then serves as a security lock. 
For larger $\gamma$ more tiles are preserved, and for smaller $\gamma$ we have more certainty that the circle will be fully tiled. 
In practice $\gamma = 0.5$ worked for all cases, although in some cases we used $\gamma = 0.7$ to speed up the simulation, especially cases where the rank was small.

\section{Computational complexity}
In this section we will analyze the computational complexity of finding the tile that contains a point $\vec{P}$ at a distance $R$ from the center of symmetry in a tiling of folding symmetry of order $N$. 
We are then interested in knowing how the computation time increases when increasing $R$ and increasing $N$ for constructing this tilling and then realizing the search.
We will focus on analyzing the algorithms based on the cut-and-projection and the generalized dual method although some of the calculations we will do for the generalized dual method also apply to the deflation/inflation method. 
The reason why we do not consider algorithms based on the deflation/inflation method is because, as far as we know, they do not have a general way of constructing the rules for generating tessellations of arbitrary symmetry, so, we can not compare the computational complexity as a function of the folding symmetry $N$. 

\begin{figure}
    \centering
     \noindent\begin{tabular}{@{\hspace{0.0em}}c@{\hspace{0.0em}}c@{\hspace{0.0em}}}
    \includegraphics[width=175pt]{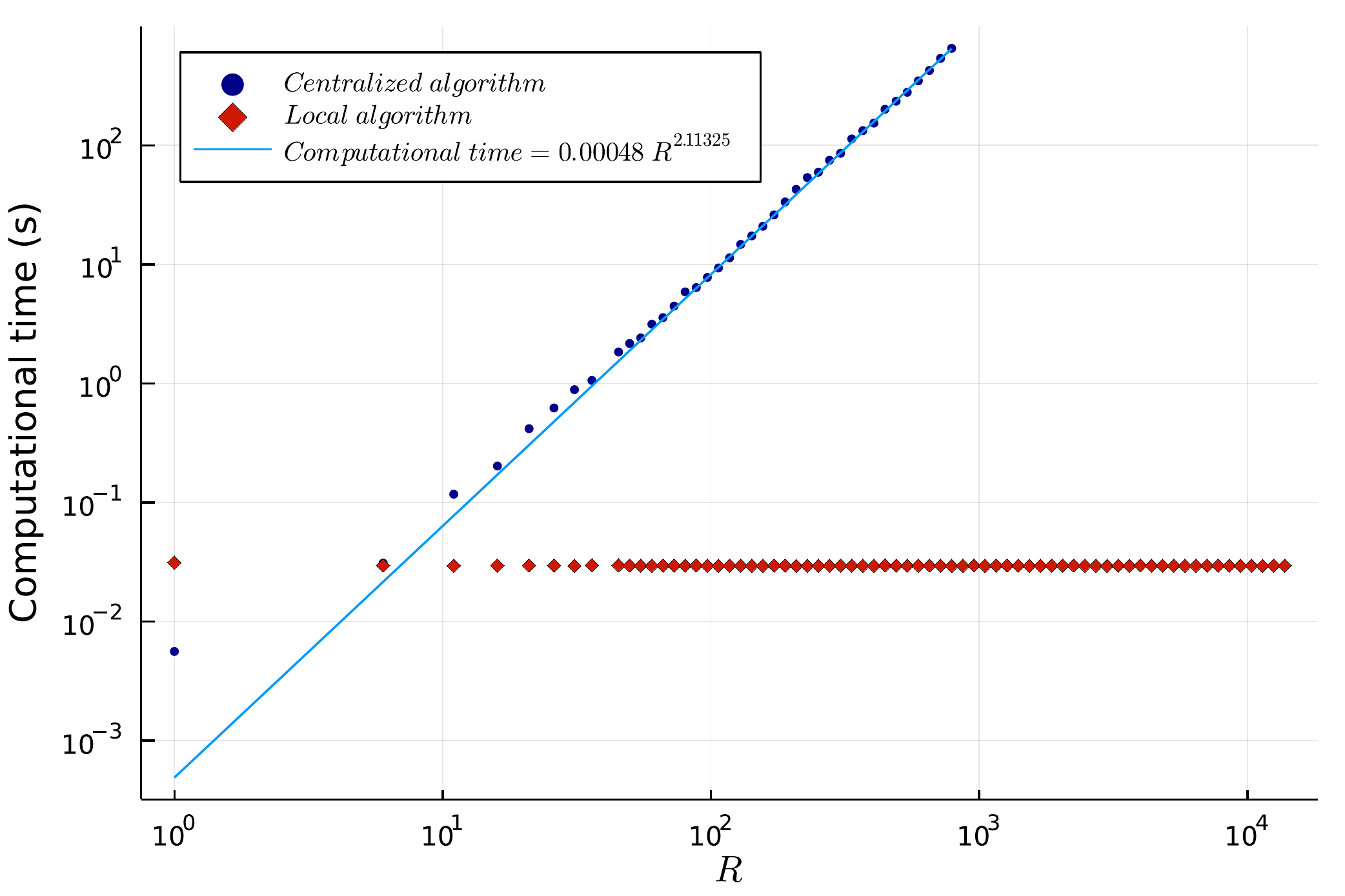} &
    \includegraphics[width=175pt]{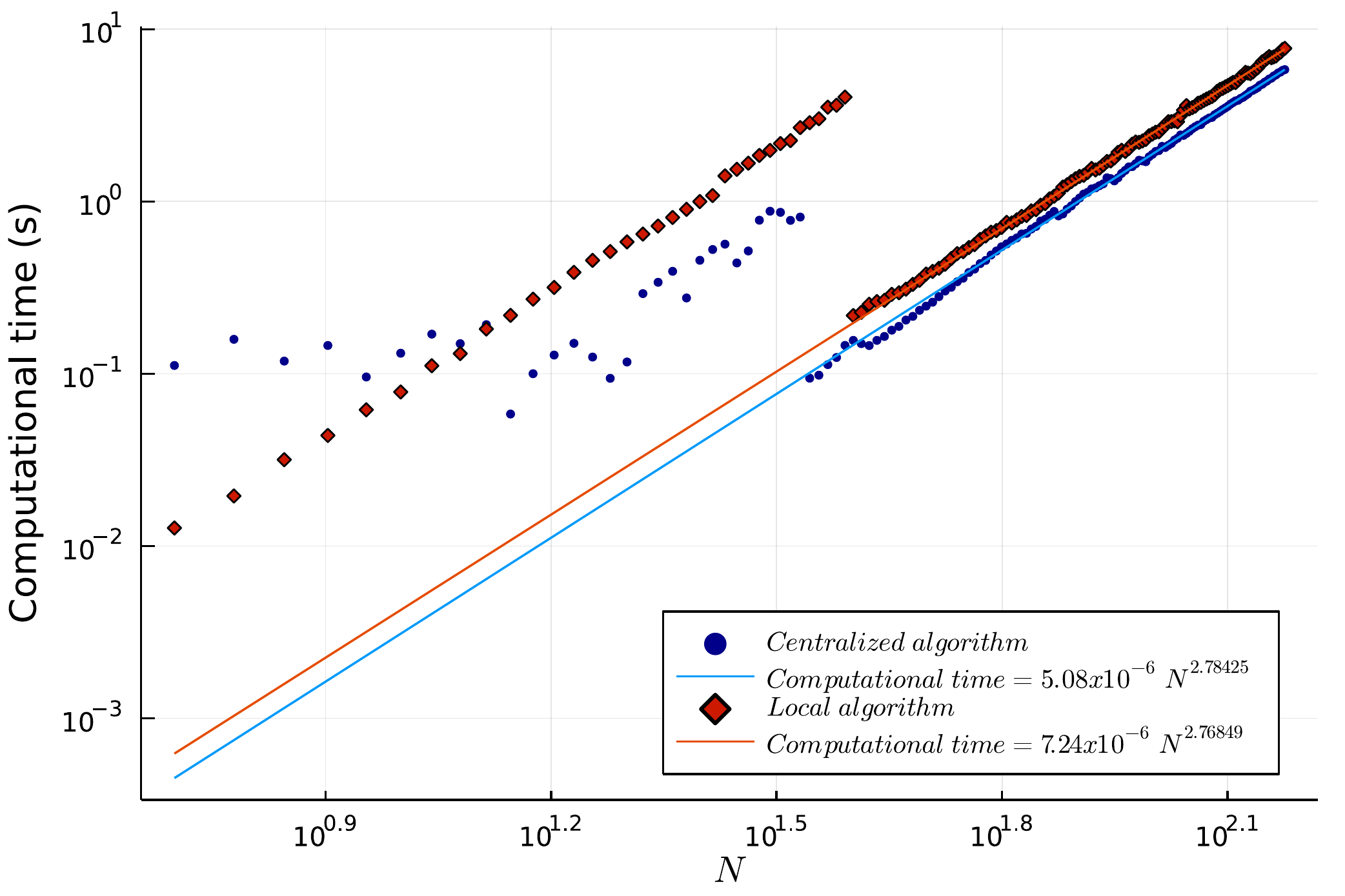}\\  
    (a) & (b)
    \end{tabular}
    \caption{(Color online) (a) Computational time vs $R$ and (b) computational time vs $N$ for the centralized and local algorithm. The adjustment parameters with their respective margin of error in parentheses are: (a) Centralized algorithm $0.00048 (8.94724 \times 10^{-5})$ and $2.11325 (0.02823)$; (b) Centralized algorithm $5.08 \times 10^{-6} (3.55 \times 10^{-7})$ and $2.78425 (0.01434)$; (b) Local algorithm $7.24 \times 10^{-6} (8.68 \times 10^{-7})$ and $2.76849 (0.02465)$}
    \label{fig: computational time 1}
\end{figure}

In the case of the classic algorithms based on the generalized dual method, to find the tile in which the particle is located, it is required to build all the tiles around the center of symmetry up to a distance $R$ (this can be done by using the explicit formulas of \cite{Naumis2003}, which is already an improvement of a brute force algorithms using the generalized dual method). 
This implies that the computational complexity to find the tile containing the particle is at least $O(R^2)$, and the memory needed to do the search is also of order $R^2$ (although a centralized (naive) algorithm does not produce exactly the tiles inside the circle of radius $R$, but many more, which is why the complexity is actually greater than $O(R^2)$, as we can see in the numeric measurements of figures \ref{fig: computational time 1} (a)). 
To this time we must add the time it takes to check the tiles to find which one contains the particle. This is in a maximum time of order $O(R^2)$, so the complexity is still dominated by $O(R^2)$. 
All this assuming that the number of tiles in a circle of radius $R$ goes as $R^2$, something that should be true since quasiperiodic tilings are hyperuniform \cite{torquato2018hyperuniform}.

In the case of the cut-and-project method first we decompose the Euclidian space $E^N$ in two $\Epar$ and $\Eperp$, where $\Epar$ is the physical space and $\Eperp$ is the orthogonal space to $\Epar$, such that $E^N = \Epar + \Eperp$, where $+$ is the direct sum. We will call $W$ the projection of a hypercube of side $1$, dimension $N$ and centered at the origin. 
We construct the quasiperiodic lattice in $\Epar$ by projecting the points of a periodic lattice in $E^N$ to $\Epar$ if they are inside a band formed by the Cartesian product of $\Epar$ and $W$. 
To obtain which are the vertices of the quasiperiodic lattice that are  neighbors to $\vec{P}$ it is not necessary to build the entire tessellation around the center of symmetry as in the generalized dual method; instead, it is only necessary to build the tessellation on the line that connects $\vec{P}$ with the center of symmetry. 
We can think it as the trajectory of a point with the initial position at the origin in the higher space that goes in direction of $\vec{P}$. 
This trajectory can be computed by applying periodic boundary conditions in a unitary cell (hypercube) until it reaches the point $\vec{P}$. 
This means that the computational complexity to produce and find which tile contains $\vec{P}$ is of order $O(R)$. 
We can improve this algorithm by noticing that the position of the point in a unitary hypercube after applying periodic boundary conditions is the fractional part of each coordinate of $\vec{P}$. 
Then we need just check which of the neighboring points of the $N-$dimensional lattice are inside the band. 
So, the time to find the tile containing $\vec{P}$ is constant (but maybe large) as a function of $R$, and the same applies to the required memory. 
The difficulty of this method is to check which points of the lattice in $E^N$ are projected. 
The first step is to obtain $W$.

The time to compute $W$ depends on $N$. 
The computational complexity when increasing symmetry $N$ of the system is then at least the time to compute $W$. 
To compute $W$ we can project all the vertices of the hypercube in $\Eperp$ and then obtain the convex hull of those points. 
This means getting the convex hull of $2^N$ points in a space of dimension $N-2$ (if $\Epar$ has dimension 2). 
The most optimal algorithm to obtain a convex hull with $n$ points in dimension $d$ has a computational complexity $O(n \log(n)+ n^{\left\lfloor d/2 \right\rfloor})$  \cite{chazelle1993optimal}. 
Substituting $n = 2^N$ and $d = N-2$ we obtain a complexity of $O(2^N log(2^N) +(2^N)^{\left\lfloor (N-2)/2 \right\rfloor}) \sim O(N 2^N + 2^{N(N-2)/2})$ which is the total computational complexity to find the tile that contain the point $\vec{P}$. 
The necessary memory to obtain the convex hull (and find the tile that contains $\vec{P}$) is at least of the order of the number of points, $2^N$. 
Therefore, the naive algorithm using cut-and-projection method is efficient only for very low symmetries. 
We did not numerically explore the computational complexity by directly applying the cut-and-project method because for symmetries above 5 the computational time grew too fast, so it was not possible to study.

The computational complexity of the generalized dual method as a function of $N$ is proportional to the number of combinations of 2 star vectors times the complexity time to sum $N-2$ integers because of equation \ref{ec:Punto_1}, i.e. $O(\frac{N(N-1)}{2})O_{sum}(N)  = O(N^2)O_{sum}(N)$, where $O_{sum}(N)$ is the complexity time for the sum of $N$ integers. 
The computational time to sum $N$ integers depends on which kind of numbers are sum, for example $N$ equal numbers can be sum in constant time, but for general sequence $O_{sum}(N)\leq O(N)$, so we expect that the computational complexity of the algorithm as a function of the symmetry is $ O(N^a)$ with $2\leq a \leq 3$. 
Because every combination of start vectors use $\sim R^2$ possible numbers ($n_j$ and $n_k$ of equation \ref{ec:Punto_1}), the total complexities is of order $O(N^a R^2)$. 

The algorithm we propose has a constant computational complexity as a function of $R$. 
As a function of $N$ it should be similar to applying naively the generalized dual method to produce a tilling around the center of symmetry, except that we also applied a Voronoi tessellation, and this cost some time (but also reduces the number of tiles where we search $\vec{P}$).

We do not know the final number of vertices we produce around $\vec{P}$, as a function of $N$, nor the number of iterations $I_V$, so we can not compute the computational complexity analytically. 
An approximation to the number of vertices can be made by considering the $\sim N^2 (1+2\beta)^2 \sim N^2$ intersections in the dual grid, which is equal to the number of rhombuses that are produced before ``cleaning'' the disconnected rhombuses from the main cluster. 
We don't know how many vertices this produces (since some rhombuses are disconnected and others share vertices), but we can overestimates this by 4 vertices for each rhombus. 
This leads to a complexity of obtaining the Voronoi polygons of $O(4N^2 log( 4N^2 )) \propto O(N^2 log( N ) )$. Finding which tile it is intakes less than $O(N^2 \beta^2)$ time. 
This means that the time to find the tile that contains the point is of the order $O(N^2 log(N))I_v$. 
Therefore, at least the complexity time as a function of $N$ will be $N^2 log(N)$. We do not know how to calculate $I_V$, so instead, we have measured numerically the computational time to find the tile that contains a fixed point $\vec{P}$ as a function of $N$ using both algorithms, the centralized algorithm, and our (local) algorithm. 

In figure \ref{fig: computational time 1} (a) we show the computational time as a function of $R$ for $N = 5$. 
We observe that the computational time grows slightly faster than $O(R^2)$ with the centralized algorithm, while our (local) algorithm keeps a constant computational time. 
We also observe that our algorithm becomes more efficient already for radii greater than $R = 10u$ where $u$ is the length of one side of the tiles. 
On the other hand, we measured the computational time as a function of $N$ for a fixed $R$. 
For the centralized algorithm we use a fixed distance at $R = 10u$, while for the local algorithm we take advantage of the constant computational time as a function of $R$ and we use a much higher $R$ (to be able to have better statistics) so that this radius was set to $R = 1000u$. 
The measured computational time is similar (slightly faster for the centralized algorithm) for both algorithms as is shown in figure \ref{fig: computational time 1} (b). 
In this case, the computational time as a function of the folding symmetry $N$ goes as $\approx N^{2.8}$, which agrees with the estimation we did in the previous paragraphs. 

\begin{figure}
    \centering
     \noindent\begin{tabular}{@{\hspace{0.0em}}c@{\hspace{0.0em}}c@{\hspace{0.0em}}}
    \includegraphics[width=175pt]{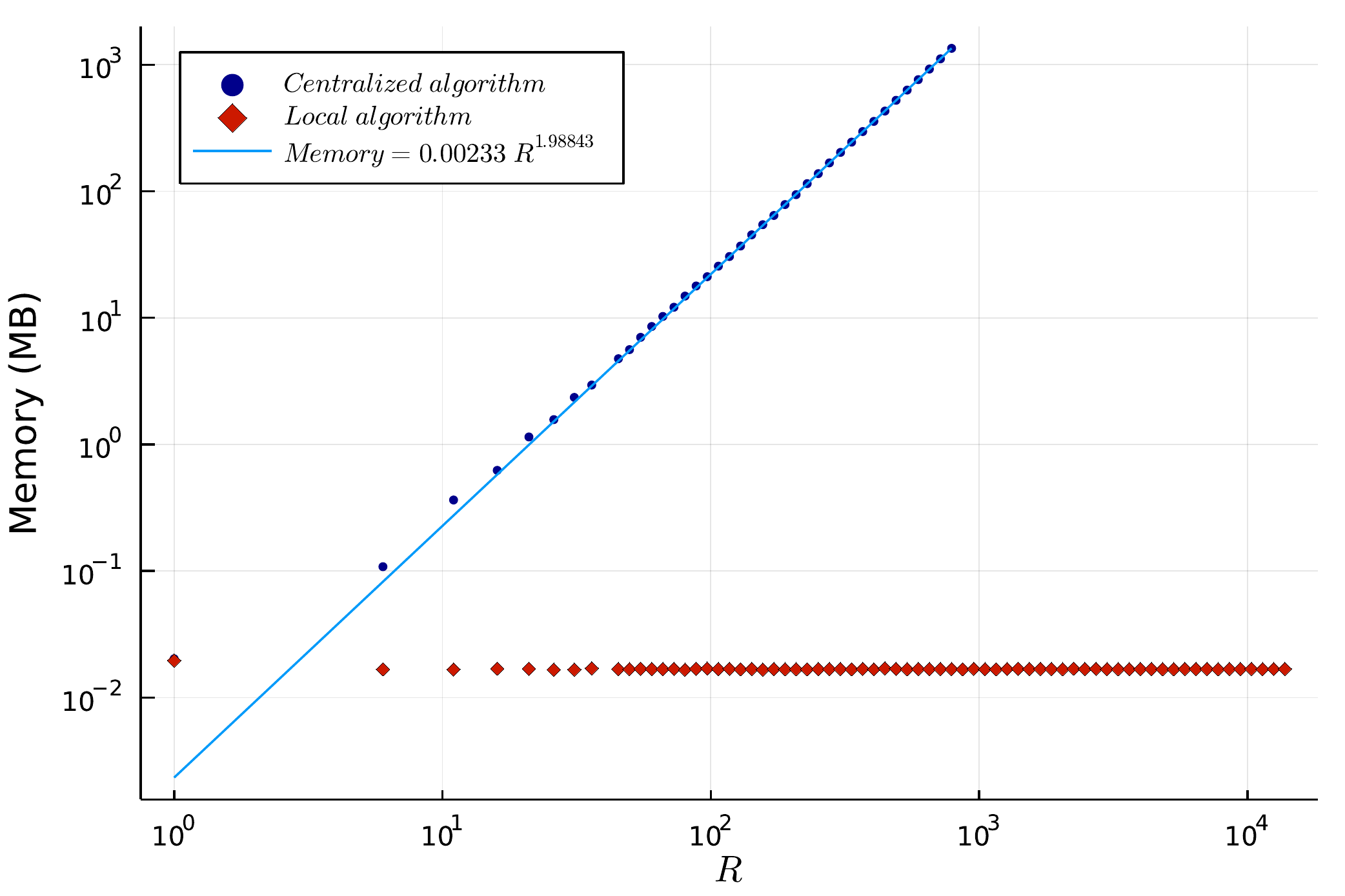} &
    \includegraphics[width=175pt]{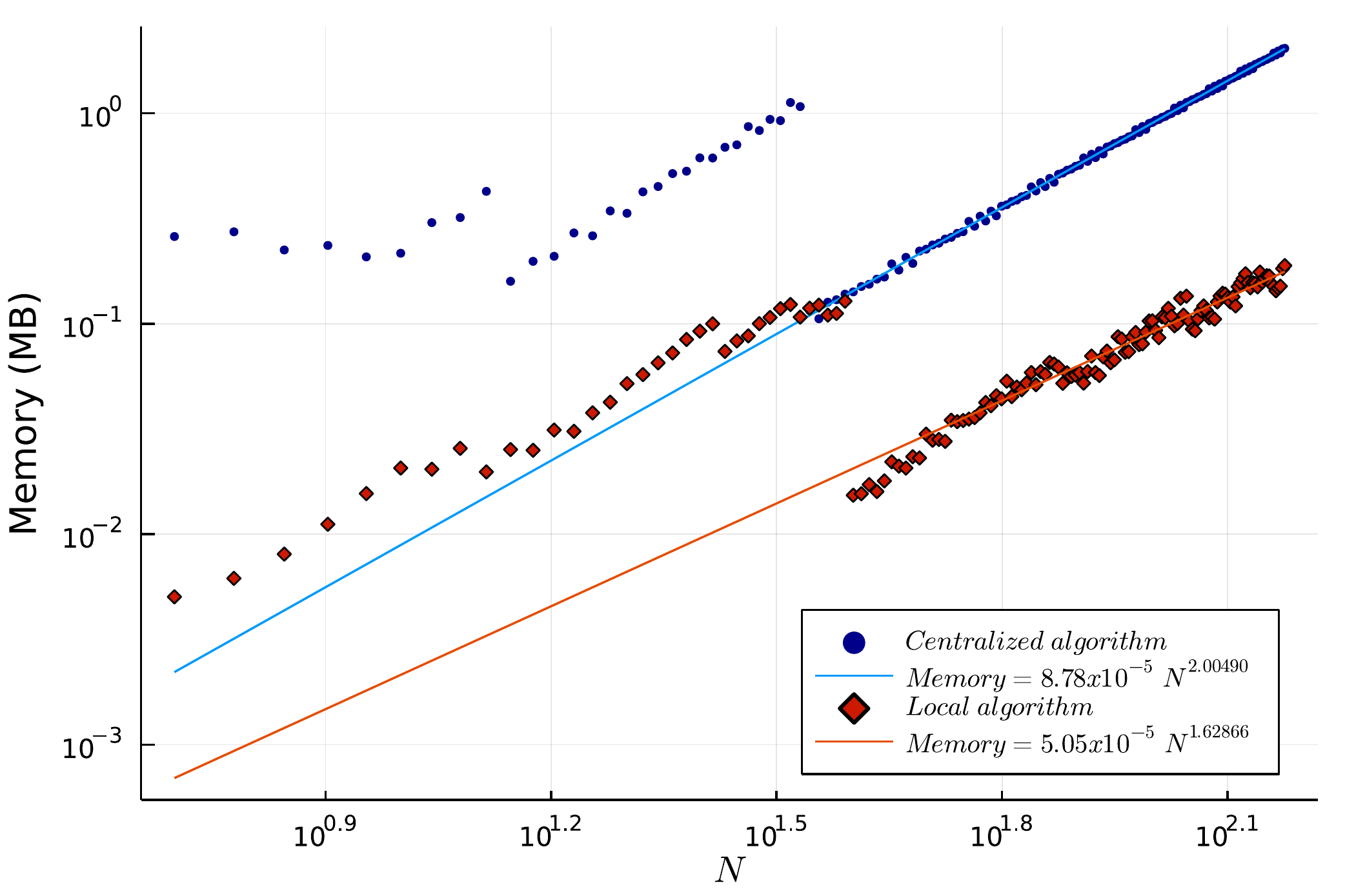}\\  
    (a) & (b)
    \end{tabular}
    \caption{(Color online) (a) Memory usage vs $R$ and (b) Memory usage vs $N$ for the centralized and local algorithm. The adjustment parameters with their respective margin of error in parentheses are: (a) Centralized algorithm $0.00233 (3.30338 \times 10^{-5})$ and $1.98843 (0.00218)$; (b) Centralized algorithm $8.78 \times 10^{-5} (6.15 \times 10^{-6})$ and $2.00490 (0.01452)$; (b) Local algorithm $5.05 \times 10^{-5} (2.08 \times 10^{-5})$ and $1.62866 (0.08611)$}
    \label{fig: computational time 2}
\end{figure}

Memory usage is another relevant quantity in the performance of simulations. 
A lower bound to memory usage is the memory used to store the vertices of the tiling. 
Figure \ref{fig: computational time 2} shows the memory consumption to store the tilling around $\vec{P}$ for both algorithms as a function of $R$ and as a function of $N$. 
We can see that the centralized algorithm grows as $R^2$ as expected, while our algorithm keeps a constant memory usage. 
As a function of $N$, memory usage grows as $N^2$ for the centralized algorithm while only as $N^{1.6}$ for our algorithm. 
The difference in both cases is the fact that our algorithm eliminates the tiles that do not belong to the main cluster, while this is not done in the other algorithm.  
The jump for the local algorithm for folding symmetry around 40 is because for smaller symmetries a margin of error $\beta = 1$ is necessary, while for larger symmetries $\beta = 0$ is enough. 
In the case of the centralized algorithm, coincidentally also close to the same symmetry, but this is because from that symmetry it is necessary to take only 1 value for each star vector ($n_j$ and $n_k$ in equation \ref{ec:Punto_1}) to produce a cluster that has a radius larger than $R = 10u$, while for smaller symmetries it is necessary a wide range of numbers. 
The jump occurs in another symmetry if we choose bigger values of $R$. 

In the literature we have found other strategies to generate symmetric quasiperiodic lattices, as \cite{reiter2002atlas} or \cite{konevtsova2011constructing}, but none of them study the computational complexity, nor do they seem to be applicable far from the center of symmetry or for high symmetries.

\section{Simulations}

Once we have a strategy to generate the vertices of a quasiperiodic lattice around any point in the plane, we can simulate the motion of a particle in a quasiperiodic environment considering the influence of some potential centered on the vertices of the lattice. 
Typically in the simulations, the tight-binding condition is used, which means only considering the closest potentials i.e. we only considered the potentials associated with the vertex of the Voronoi polygon (which we already obtained in the process of construction of the main cluster) in which the particle is found. 
Therefore, once you have the Voronoi cells, the motion of the particle can be simulated if you have a function that, given the Voronoi cell where the particle is entering, the position and speed with which it enters, gives the position and velocity of the particle when it leaves the Voronoi cell (and therefore the new cell it enters) and the time taken by the particle in this process. 
In general, the restriction of first neighbors is not necessary, for example, all Voronoi cells that are neighbors to the cell where the particle is located can also be considered (second neighbors approximation) to determine where said particle enters and leaves. 

The steps for simulating a particle in a quasiperiodic environment for a given time are: 
(i) A quasiperiodic lattice is generated around the initial position of the particle, saving the Voronoi polygons associated with the vertices of the lattice. 
(ii) It is located in which Voronoi polygon the particle is. 
(iii) Given a potential centered in the center of the polygon, the trajectory of the particle is obtained until it leaves the polygon and enters another.
(iv) Step (iii) is repeated until the particle reaches the cluster boundary, in which case step (i) is repeated or until the desired simulation time has passed, in which case the simulation stops. 

\subsection{Quasiperiodic Lorentz gas}

As examples of the use of this algorithm, we will study quasiperiodic Lorentz gases, reproducing the results in \cite{kraemer2015horizons} to test the algorithm, and in the Boltzmann-Grad limit for different symmetries. 
A Lorentz gas consists of a set of fixed obstacles centered at the vertices of a lattice and a particle that moves freely until it collides with some obstacle; in that case it typically follows a specular reflection. 
In the case of a 2D quasiperiodic Lorentz gas, the lattice is a quasiperiodic lattice and the obstacles are typically disks. 

\subsubsection{Locally finite horizons}

Lorentz gases can be classified in terms of their horizons, i.e. if particles can have free flights of infinite length or not. In this case, there are 3 possibilities: finite horizon, if the length of free flights has an upper bound, infinite horizon if there are flights of infinite length and locally finite horizon if there are no flights of infinite length but there is not an upper bound. 
In \cite{kraemer2015horizons}, it is shown that quasiperiodic Lorentz gases can have all 3 types of horizons depending on the size of the obstacles. 
In addition, heuristic arguments and numerical simulations were performed to show that in the locally finite regime the free path distribution must approximate a power-law with exponent $-5$. 
Here we reproduce the results, studying an equivalent Lorentz gas in the 3 regimens: finite, infinite and locally finite horizons.  

\begin{figure}
    \centering
    \includegraphics[width=350pt]{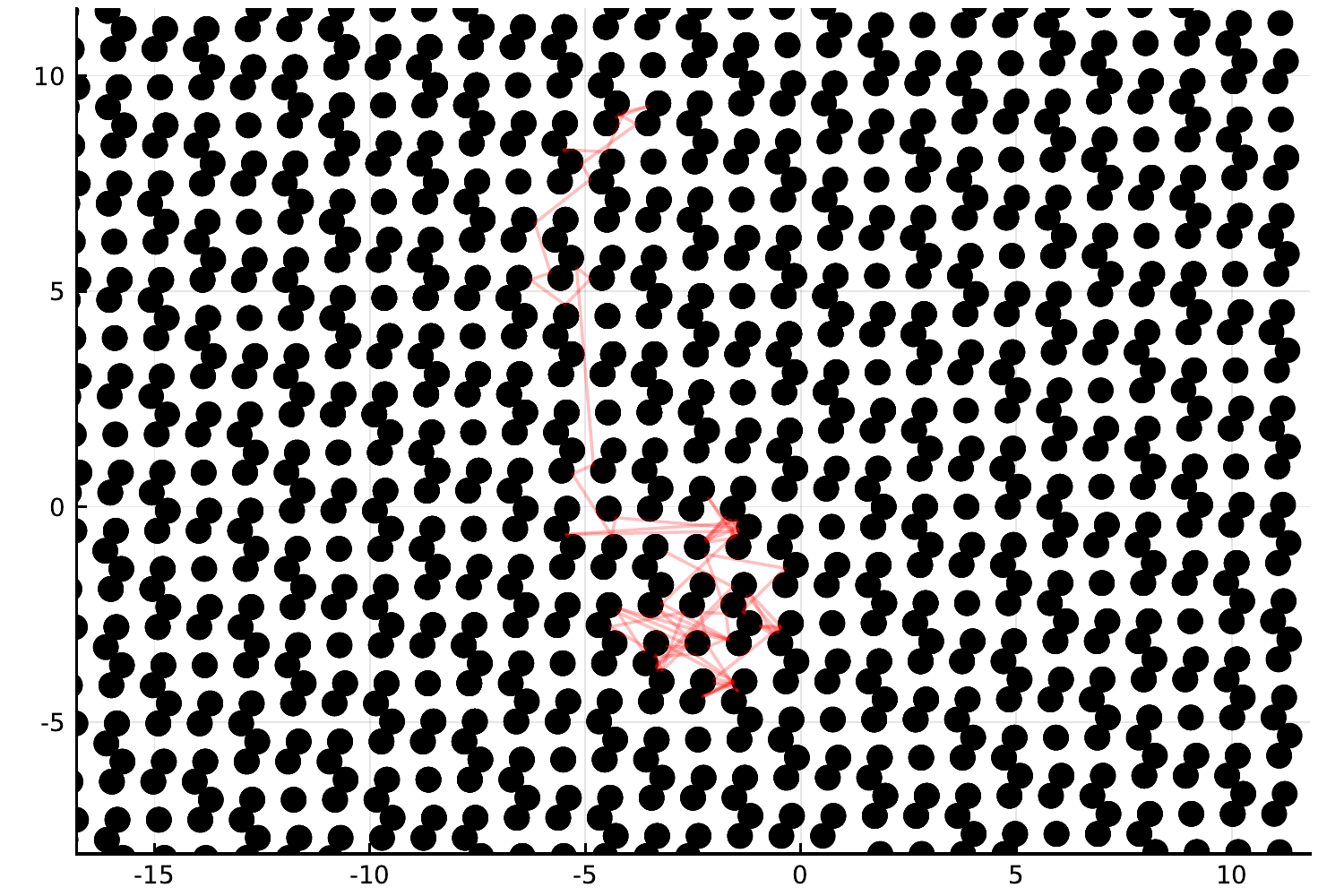}
    \caption{Trajectory in the quasiperiodic Lorentz gas similar to the array in \cite{kraemer2015horizons}}
    \label{fig: quasiperiodic3D}
\end{figure}

In \cite{kraemer2015horizons}, the quasiperiodic Lorentz gas is produced by the cut-and-projection method, projecting from a 3-dimensional space, to a 2-dimensional one. To make the projection, we can think that the lattice is generated by $N$ generating vectors $\vec{v}_i$ and the subspace where they are projected is the subspace generated by $d$ vectors of the canonical base of $\mathbb{R}^N$, in this case, the plane $xy$. 
In the GDM, equivalently, $N$ star vectors are required to generate the tessellation. The directions of the star vectors are orthogonal to the directions of the projections of the $\vec{v}_i$'s \cite{gahler1986equivalence}. 
Using directly the projections of the $\vec{v}_i$'s as a star vector will produce a rotated system and re-scaled by some constant factor. So, to reproduce the results of \cite{kraemer2015horizons} we use $(0.12861712428405658, -0.8851314226378474)$, $(-0.961044638791549, 0.0)$ and $(0.24464430831499137, 0.46534112719498627)$ as star vectors, which are the projections of the 3 generating vectors used to produce the lattice in \cite{kraemer2015horizons}. 
Figure \ref{fig: quasiperiodic3D} shows a trajectory in the resulting quasiperiodic Lorentz gas which has the same pattern as figure 1(d) in \cite{kraemer2015horizons} except that is rotated and re-scaled. 

\begin{figure}
    \centering
    \includegraphics[width=350pt]{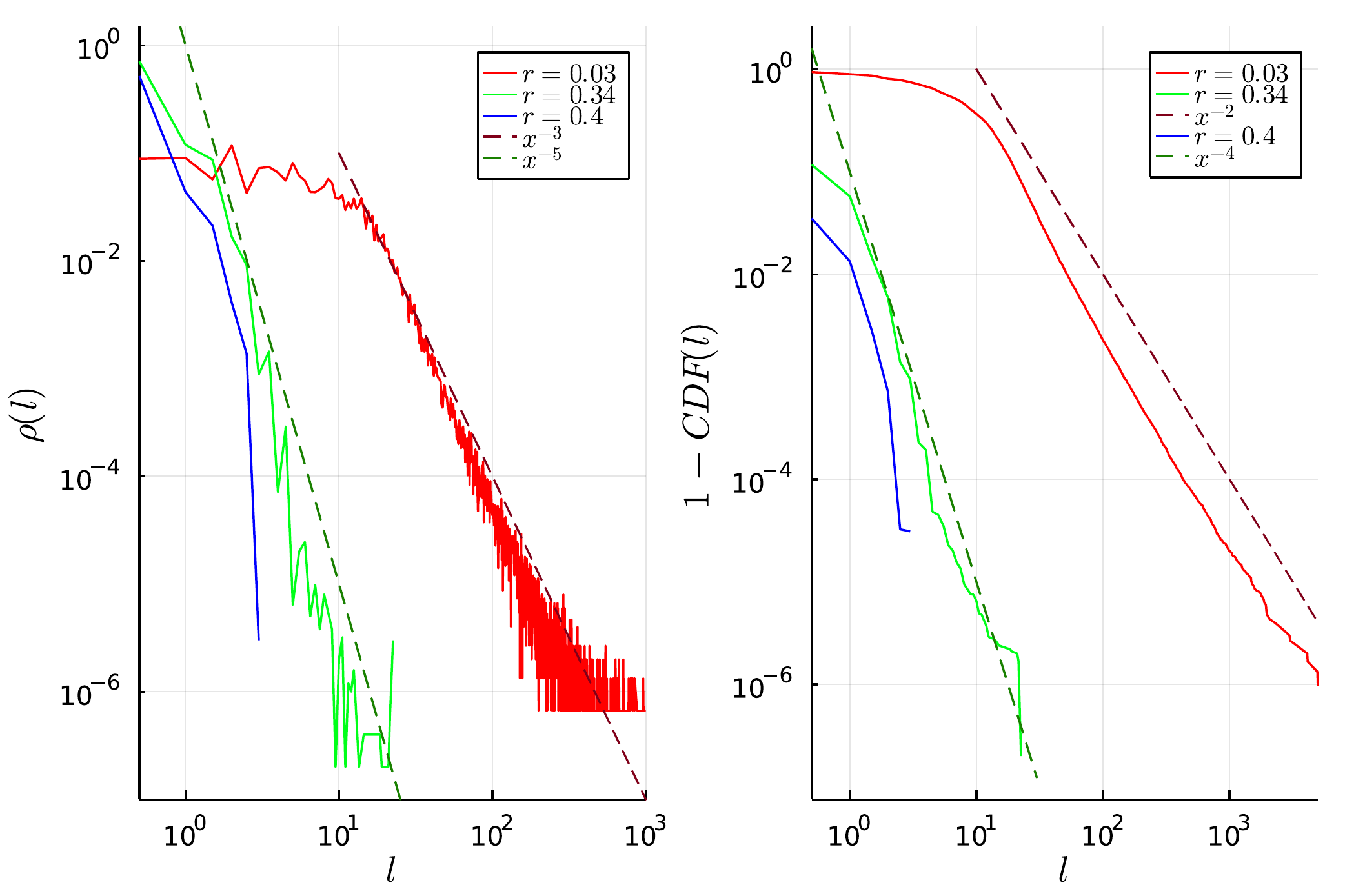}
    \caption{Free path distribution $\rho(l)$ (left) and $1-CDF(l)$ (right) as a function of $l$ for several radii, $r = 0.03$ (red), $0.34$ (green) and $0.4$ (blue).  Dotted lines are power-laws with exponents $-3$ (dark red) and $-5$ (dark green) on the left and $-2$ (dark red) and $-4$ (dark green) on the right.}
    \label{fig: freepath-distribution-3D}
\end{figure}

Because of the re-scaling, the critical radius where the system becomes locally finite changes, but the behavior should be the same in each regimen. 
To estimate the scale factor $\kappa$ we first found numerically the critical radius, i.e. the radius at which $1-CDF(l)$ is a power-law-like function with exponent $-4$ (or equivalently where the exponent of $\rho(l)$ is $-5$), where $CDF(l) = \int_0^l \rho(l)$ is the cumulative density function. 
Note that if $\rho(l) = exp(-\lambda l)$ then $1-CDF(l) \propto exp(-\lambda l)$, and if $\rho(l) = a l^{-b}$ then $1-CDF(l) \propto l^{b-1}$. 
For this purpose we perform simulations to obtain $10^6$ free flights for radii in the range $[0.01,0.5]$ separating each radius by $0.01$. 
Once we have a candidate to a critical radius, we increase the number of flights in those radii to $10^7$. 
We obtained a critical radius of $r_c \sim 0.34$. 
Then we divided it by the critical radius calculated in \cite{kraemer2015horizons} ($r \sim 0.309$) to obtain $\kappa \sim 1.10032362$). 
Figure \ref{fig: freepath-distribution-3D} shows the free path distribution $\rho(l)$ and the function $1-CDF(l)$ for $r = 0.03, 0.34$ and $0.4$. 
For $r = 0.03$ we obtained $3\times 10^6$ free flights, while for $r = 0.34$ and $r = 0.4$ we obtained $10^7$ free flights. 
These 3 radii are the closest radii we simulate equivalent to the radii used in \cite{kraemer2015horizons}. 
As a visual aid, we add (dotted lines) power-laws with exponents $-3$ and $-5$ on the left and with exponents $-2$ and $-4$ on the right.
We can see that these results are practically identical to those observed in \cite{kraemer2015horizons}

\subsubsection{Boltzmann-Grad limit}

The Boltzmann-Grad limit consists of reducing the radius of the obstacles at the same time that the length of the free paths is re-scaled by the diameter of the disks.

When the quasiperiodic Lorentz gas is not symmetric, it has been shown that corridors where particles can lay out infinite free paths exist \cite{kraemer2013embedding}, similar to the periodic case. As a result a similar distribution of free paths to that of the periodic Lorentz gas was found for radii small enough \cite{kraemer2015horizons} as we have seen in the previous example. These corridors can be blocked by growing the obstacles until a critical radius, but for quasiperiodic Lorentz gases with rotational symmetry they maybe only exist finite and locally finite horizons \cite{kraemer2013embedding} in which case the distribution is unknown \cite{marklof2014free}.  

The free path lengths distribution in the Boltzmann-Grad limit is of interest not only because of its mathematical importance (and in the solution of the Boltzmann kinetic equation) but also, for the periodic case, it is related to the distribution of energy gaps in 2D quantum harmonic oscillator \cite{Marklof2021QuantumOscilators}. 
However, although the periodic case has been extensively studied \cite{marklof2011periodic, marklof2008kinetic, kraemer2015efficient, caglioti2010boltzmann, golse2006periodic, caglioti_golse, golse2012recent, griffin2021quantum}, the quasiperiodic system has been studied relatively little \cite{marklof2014free, kraemer2015horizons, wennberg2012free} despite its importance; among other reasons because of the difficulty of doing numerical simulations \cite{wennberg2012free, kraemer2013embedding} where the use of approximants is avoided. 

Considering that a quasiperiodic Lorentz gases that can be produced with the cut-and-project method can be seen as particles moving in a higher-dimensional space with periodically placed obstacles, one might expect the behavior of the free path length distribution to resemble that of a periodic Lorentz gas \cite{marklof2011periodic}. 
In which case the behavior of the tail of the distribution would be that of a power-law with exponent $-3$, and said tail would dominate from probabilities less than $\frac{2^{2-d}} {d(d+1)\zeta(d)}$ (Theorem 1.14 of \cite{marklof2011periodic}), where $d$ is the dimension and $\zeta(d)$ is the Riemann zeta function, which is approximately $1$ for $d>>2$. 
However there are some differences between the quasiperiodic and the periodic Lorentz gas as we explain below. 
In the periodic case, the free path distribution can be calculated by obtaining the probability that a $d-$dimensional cylinder (pointing in any direction) of length $l$ (the length of the free path) and radius $r$ contains an integer coordinate point. Which is equivalent to the probability that a vertex of a periodic lattice (rotated randomly) intersect a cylinder laying in the $x$ axis. 
In the quasiperiodic case, each trajectory is constrained to move in a completely irrational plane of the $N-$dimensional space, which is equivalent to moving in a periodic cell in the higher dimensional space with periodic boundary conditions \cite{kraemer2013embedding}. 
So, the cylinders are not pointing in any direction, but they cover the periodic cell. 
Another difference is that instead of measuring the probability that a point of integer coordinate is inside the cylinder, we should consider a $(N-d)-$polytope centered at the integer coordinate and orthogonal to the plane of motion. 
This polytope is the projection of the unitary hypercube in the orthogonal space, which is bounded at the inside and outside by two balls. 
The greater the dimension of the hypercube is, the more the radius of those balls is closer, so for $N>>1$ the polytopes can be approximate by spheres of dimension $N-d$. 
The question is to calculate the probability that a cylinder of length $l$ is crossing a sphere of dimension $N-d$ centered in integer coordinates. 
Because the plane of motion is completely irrational and the spheres are ``flat'' (of dimension $N-d$), we expect a similar behavior despite the differences in both cases. 
If the distribution of free paths tends to a power-law with exponent $-3$ as in the periodic case, we can think in $\frac{2^{2-d}} {d(d+1)\zeta(d)}$ as an upper bound to the probability the distribution is dominated by the power-law. 
The reason is that the power-law is the result of the existence of channels (or principal horizons) \cite{dettmann2012new}. 
The width of the channel (of dimension $d-1$ or $r-1$ for quasiperiodic systems) is what delimits this probability. 
A smaller width of the channel means a smaller probability from which it dominates power-law behavior. 
As in quasiperiodic arrays the points are replaced by objects of dimension $r-2$, the width of these channels can only be reduced and therefore the probability from which the power-law behavior becomes dominant.

Furthermore, in the limit where $d$ tends to infinity, the tail of the free path length distribution for the periodic case tends to an exponential distribution according to \cite{marklof2000point}. 
So we expect this behavior when $N$ grows in the quasiperiodic if for low $r$ we find a power-law of exponent $-3$. 

On the other hand, all these results are for long flights, so it does not tell us anything about the behavior of the distribution for short paths.

In the 2D periodic case, for short flights, there is a constant distribution \cite{marklof2008kinetic,Marklof2021QuantumOscilators}, something that is not maintained when increasing the dimension of the Lorentz gas. 
As far as we know, there is no analytic results for short flights for other dimensions than $2$. 
To obtain what kind of distribution appears for probabilities greater than $\frac{2^{2-d}} {d(d+1)\zeta(d)}$, it is then necessary to do simulations. 
Up to this point, we have compared quasiperidic Lorentz gases with periodic Lorentz gases in higher dimensions using 2 quantities to refer to the dimension of the system, that of the periodic arrangements and that of the cut-and-project method. 
In \cite{reiter2002atlas} it is shown that quasicrystals with rotational symmetry $N$ can be produced by the projection of a periodic lattice of dimension $N$, and this may make us think that the equivalent dimension of the quasiperiodic systems is $N$. 
However, the cut-and-project method may use a dimension greater than the minimum required to produce the quasiperiodic lattice. 
This minimum dimension necessary to produce the quasiperiodic lattice is known as the rank $r$. 
The rank $r$ is the equivalent dimension to that of the periodic Lorentz gas in high dimensions and not the symmetry $N$. In other words, we expect similar behaviors for quasiperiodic Lorentz gases with similar ranks, and so, for probabilities less than $\frac{2^{2-r}} {r(r+1)\zeta(r)}$ we would expect the power-law behavior to dominate in the distribution of free flights. 

As we mentioned before, we can calculate $r$ from the symmetry of the quasiperiodic array using Euler’s totient function. 
Then, for quasiperiodic Lorentz gases of rotational symmetry of order $5$, $8$ and $12$ we would expect the same behavior in the tail of their distributions, where the power-law should appear from producing $2^{4-2} 4 \cdot 5 = 80$ free flights. 
However, for rotational symmetry of order $7$, where $\phi(7) = 6$, we expect the power-law to begin to appear from $2^{6-2} 6 \cdot 7 \approx 700$ free flights, and for those with symmetry of order $13$ we hope that the power-law begins to be seen from $2^{12-2} 12 \cdot 13 \approx 160,000$ free flights. 
Thus, despite having a similar symmetry in the quasiperiodic arrangements with rotational symmetry $7$ and $8$ or $12$ and $13$, we expect noticeably different behavior in their distributions. 

To do the simulations, the function that was used to obtain the trajectory inside a Voronoi cell was simply a straight line if it did not intersect an obstacle, and two segments following a specular collision with the obstacle otherwise.
Similar to what is typically done in the periodic Lorentz gas \cite{kraemer2015efficient} 

\begin{figure}
    \centering
    
    \includegraphics[width=350pt]{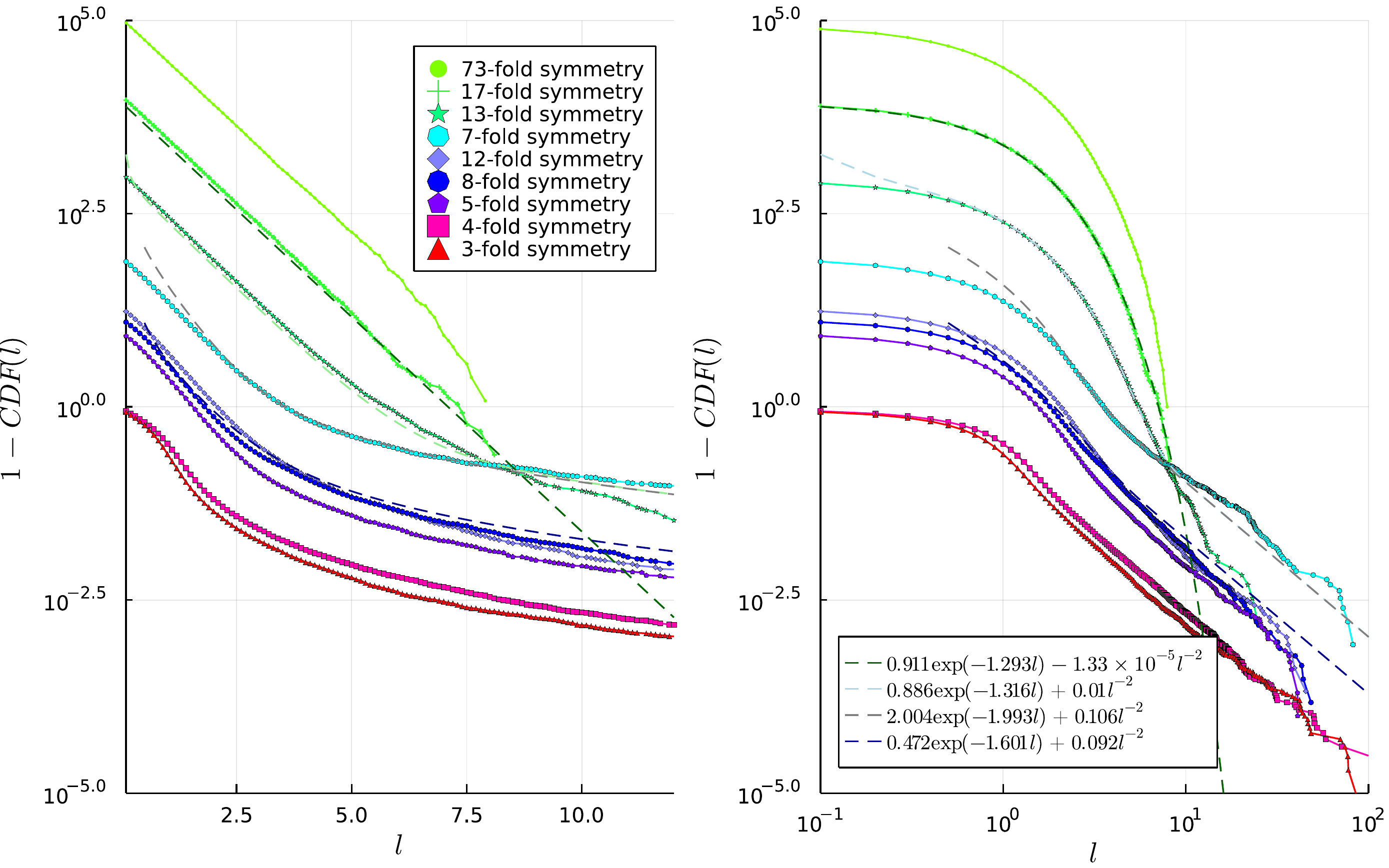}
   
    \caption{(Color online) $1-CDF(l)$ as a function of $l$ for different symmetries. The graphs in red are the $1-CDF(l)$, while the ones above are the $1-CDF(l)$  multiplied by $10$, $15$, $21$, $10^2$, $10^3$, $10^4$, and $10^5$ respectively to better visualize the data. On the left $1-CDF(l)$ is plotted with a logarithmic scale for the $y$ axis. On the right both axes are on a logarithmic scale. In this case, it can be seen that for low ranks a power-law is obtained in the tails of the distribution.}
    \label{fig:Distribucion_vuelos_libres}
\end{figure}

We perform simulations on arrays with rotational symmetry of order $3$, $4$, $5$, $7$, $12$, $13$, $17$ and $73$, obtaining $10^5$ free paths for each symmetry, except for the rotational symmetry $13$ where we have obtained $6\times 10^5$ free flights (to see the deviation to the exponential decay in the free path length distribution). 
Because the free path distribution $\rho(l)$ has a lot of noise, and so, it is difficult to visualize several plots together, we used instead of the cumulative density function $CDF(l)$. 
In the left side of figure \ref{fig:Distribucion_vuelos_libres} we show $1-CDF(l)$ in logarithmic scale for the $y$ axis and on the right side the same distribution in logarithmic scale in both axes. 
We fitted the data using $a_1 exp(a_2 l) + a_3 l^{-2}$ as the fitted curve with 3 parameters, and the obtained equations are shown in the right side of the figure. 
The obtained parameters are: $a_1 = 0.472 \pm 0.06471$, $a_2 = 1.601 \pm 0.05247$ and $a_3 = 0.092 \pm 0.00088$ for rotational symmetry $5$,  
$a_1 = 2.004 \pm 0.64065$, $a_2 = 1.993 \pm 0.11936$ and $a_3 = 0.106 \pm 0.0019$ for rotational symmetry $7$,  
$a_1 = 0.886 \pm 0.01388$, $a_2 = 1.316 \pm 0.00624$ and $a_3 = 0.01 \pm 0.00064$ for rotational symmetry $13$ and  
$a_1 = 0.911 \pm 0.03754$, $a_2 = 1.293 \pm 0.01725$ and $a_3 = -1.33\times 10^{-5} \pm  0.00221$ for rotational symmetry $17$. 
Note that the in general $a_2$ and $a_3$ parameters reduce when the rank increases, so the exponential behavior becomes dominant. Indeed, for the most part of the data of ranks greater or equal to $12$ (symmetries $13$, $17$ and $73$), the data are well fitted only with an exponential function. 
We conclude that increasing the rank increase the region where the exponential distribution dominates, so that the complete distribution tends to an exponential distribution at the limit where $r \rightarrow \infty$, similar to the free path length distribution of a disordered Lorentz gas with a Poisson distribution of obstacles. 
This is contrary to what had previously been observed numerically in quasiperiodic Lorentz gases at the Boltzmann-Grad limit \cite{wennberg2012free}, and it might seem contrary to the result of \cite{marklof2014free}, where any quasiperiodic array of obstacles is shown to have a non-exponential distribution in the limit $l \rightarrow \infty$. 
However, it is consistent, since for any rank $r \ne \infty$ we expect the distribution to be a power-law with exponent $-3$ in the limit $l \rightarrow \infty$, but the value of $l$ beyond which the power-law behavior becomes dominant, grows rapidly with $r$.
With this result, a question that remains open is: 
In the limit of high symmetries, what kind of distribution of vertices does a quasiperiodic lattice have? Is it similar to a Poisson distribution? or maybe similar to a jammed state? \cite{torquato2010jammed, torquato2018hyperuniform}. 

\section{Conclusions}
\label{conclusions}

We have introduced an efficient algorithm to simulate particles in 2D quasiperiodic environments of any symmetry, with the advantage over other algorithms that it is not necessary to previously store the information about the vertices of the quasiperiodic lattice, but rather the algorithm includes a way to build an environment locally, which allows very long simulations without the need to use periodic boundary conditions. 
The code of this algorithm has been written in JULIA \cite{bezanson2017julia}, and can be download from
\url{https://github.com/AlanRodrigoMendozaSosa/Quasiperiodic-Tiles}. 
We tested the algorithm on quasiperiodic Lorentz gases in a locally finite horizon recuperating the results in \cite{kraemer2015horizons} and in the Boltzmann-Grad limit, obtaining the free path length distribution for several systems with different symmetries. 
It is noticeable that the distribution obtained seems to tend towards an exponential distribution when the rank of the array grows. 
As possible future applications, this algorithm could be applied to study photonic quasicrystals as in \cite{rousseau2017ray}, heat transport in quasicrystals as in \cite{larralde2003transport}, or to study the structure for high symmetries using tools as in \cite{torquato2010jammed, torquato2018hyperuniform}. It can also be useful to study Brownian Dynamics as in \cite{Martinsons2019}. 

The algorithm is based on the generalized dual method, which can be easily generalized to more dimensions, with the only difficulty of obtaining the Voronoi polyhedra in higher dimensions. 
On the other hand, the generalized dual method allows the production of quasicrystals that cannot be produced via the cut-and-project method if an aperiodic spacing is added to the grid. 
In this case, our algorithm may or may not be functional, since we are not certain that there is a bound and an average in the separation of the bands mentioned in figure \ref{fig:Stripes}. 
If it were the case, the algorithm also could be extended to this class of quasiperiodic arrangements.

\section{Acknowledgements}  
We thank Marco Lenci and especially Jens Marklof for useful discussions and the references provided to explain the obtained free path distributions. 
The authors appreciate the computing platform provided by the Laboratorio de Cómputo de Alto Rendimiento, under coordination of Departamento de Matemáticas of Facultad de Ciencias, UNAM
Financial is acknowledged from CONACYT for ARMS’s doctoral studentship and from DGAPA-UNAM PAPIIT grant IA106618. 

\vspace*{10pt}

\bibliographystyle{unsrt}
\bibliography{efficient_quasiperiodic}

\end{document}